 \pdfoutput=1
 \documentclass[journal,onecolumn,12pt, draftclsnofoot]{IEEEtran}
\IEEEoverridecommandlockouts
\usepackage{cite}
\usepackage{amsmath,amssymb,amsfonts}
\usepackage[noend]{algpseudocode}

\usepackage{threeparttable}
\usepackage{subfigure}
\usepackage[ruled,linesnumbered]{algorithm2e}
\usepackage{graphicx}
\usepackage{textcomp}
\usepackage{xcolor}
 \usepackage{makecell}
\usepackage{pdfpages}
\usepackage{algpseudocode}
\usepackage{footnote}
\usepackage{diagbox}
\usepackage{bbm}
\usepackage{dsfont}

\newtheorem{theorem}{Theorem}

\newtheorem{Remark}{Remark}
\newtheorem{Corollary}{Corollary}
\newtheorem{Definition}{Definition}
\newtheorem{Example}{Example}
\newtheorem{lemma}{Lemma}
\usepackage{comment}
\usepackage{booktabs,color,amssymb,amsmath}

\def\BibTeX{{\rm B\kern-.05em{\sc i\kern-.025em b}\kern-.08em
    T\kern-.1667em\lower.7ex\hbox{E}\kern-.125emX}}

\begin{document}

\title{ {Coded Distributed Computing for Hierarchical Multi-task Learning}
	
\author{
  \IEEEauthorblockN{ Haoyang Hu, \IEEEmembership{Student Member,~IEEE}, Songze Li, \IEEEmembership{Member,~IEEE,} Minquan Cheng, \IEEEmembership{Member,~IEEE,} and Youlong Wu, \IEEEmembership{Member,~IEEE} \\}}

\thanks{Haoyang Hu and Youlong Wu are with the School of Information Science and Technology, ShanghaiTech University, Shanghai 201210, China. (e-mail: \{huhy, wuyl1\}@shanghaitech.edu.cn). 

Songze Li is with the Thrust of Internet of Things, The Hong Kong University of Science and Technology (Guangzhou), Guangzhou, China, and also with the Department of Computer Science and Engineering, The Hong Kong University of Science and Technology, Hong Kong SAR, China (e-mail: songzeli@ust.hk). 

Minquan Cheng is with Guangxi Key Lab of Multi-source Information Mining \& Security, Guangxi Normal University, Guilin 541004, China (e-mail: chengqinshi@hotmail.com).
}

}
\maketitle
%-----------------------------------------------------------------------------------------------------
\begin{abstract}
In this paper, we consider a hierarchical distributed multi-task learning (MTL) system where distributed users wish to jointly learn different models orchestrated by a central server with the help of a layer of multiple relays.  Since the users need to download different learning models in the downlink transmission,  the distributed MTL suffers more severely from the communication bottleneck compared to the single-task learning system.  To address this issue, we propose a coded hierarchical MTL scheme that exploits the connection topology and introduces coding techniques to reduce communication loads.  It is shown that the proposed scheme can significantly reduce the communication loads both in the uplink and downlink transmissions between relays and the server.  Moreover, we provide information-theoretic lower bounds on the optimal uplink and downlink communication loads, and prove that the gaps between achievable upper bounds and lower bounds are within the minimum number of connected users among all relays. In particular, when the network connection topology can be delicately designed, the proposed scheme can achieve the information-theoretic optimal communication loads.
Experiments on real datasets show that our proposed scheme can  reduce the overall training time by 17\% $\sim$ 26\% compared to the conventional uncoded scheme.
\end{abstract}


%-----------------------------------------------------------------------------------------------------
\section{Introduction}
The development of the Internet of Things (IoT) has brought about an explosion of data, making distributed learning receive significant attention these days \cite{letaief2021edge}. The local data across distributed users are often not independent and identically distributed (Non-IID), which results in a single global model failing to capture the characteristics of the data well.
Multi-task learning (MTL) \cite{zhang2021survey,ruder2017overview,tan2022towards} is a learning paradigm that helps to exploit the non-IID property to achieve better generalization performance than learning the tasks independently by leveraging useful information contained in related tasks. In \cite{liu2017distributed, jaggi2014communication, smith2017federated, dinh2021new}, distributed MTL has been studied in which distributed users want to learn models simultaneously under the orchestral of a central server and leverage the correlation between tasks to train better-personalized models for each user.

However, the exchange of model parameters between distributed nodes incurs a huge amount of communication load, causing a communication bottleneck that limits the performance of distributed learning systems \cite{shi2020communication}. The communication bottleneck is more  severe in the distributed MTL setting. For example, in the conventional MTL  framework \cite{jaggi2014communication, smith2017federated, dinh2021new}, distributed users first perform the local update and then send generated intermediate values (IVs)\footnote{Under different distributed optimization algorithms, IVs represent local models, gradients, etc.} to
the central server via the uplink. After receiving IVs from all users, the server performs the global update phase to obtain multiple global models, and then sends each user its model separately via the downlink, so that the downlink communication load grows linearly  with the number of users. Hence distributed MTL suffers from a communication bottleneck both in the uplink and downlink.

\begin{figure}[htbp]
	\centering
        \subfigure[Uplink Communication]{
            \centering
		\includegraphics[scale=0.35]{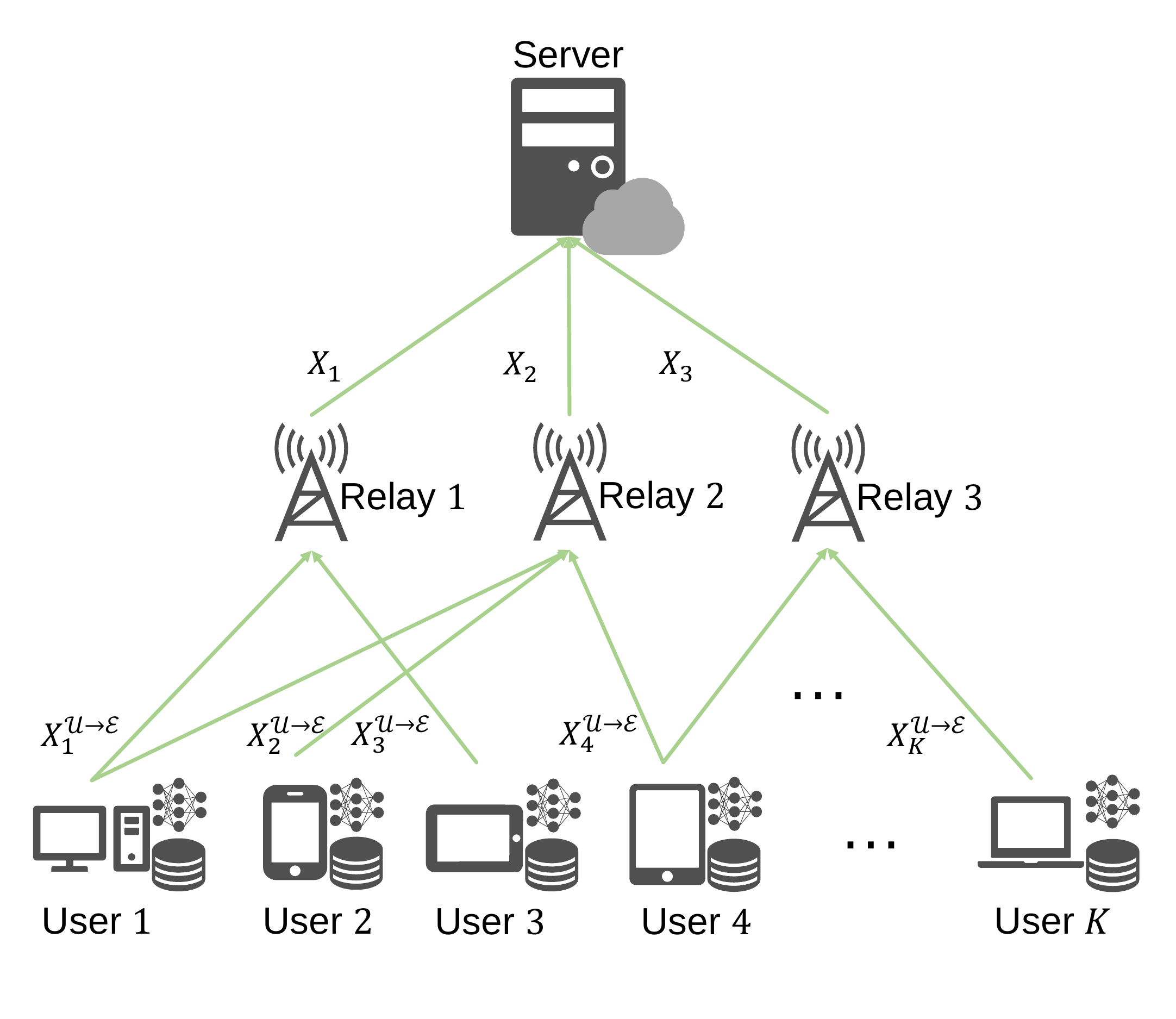}
        }
        \subfigure[Downlink Communication]{
            \centering
		\includegraphics[scale=0.35]{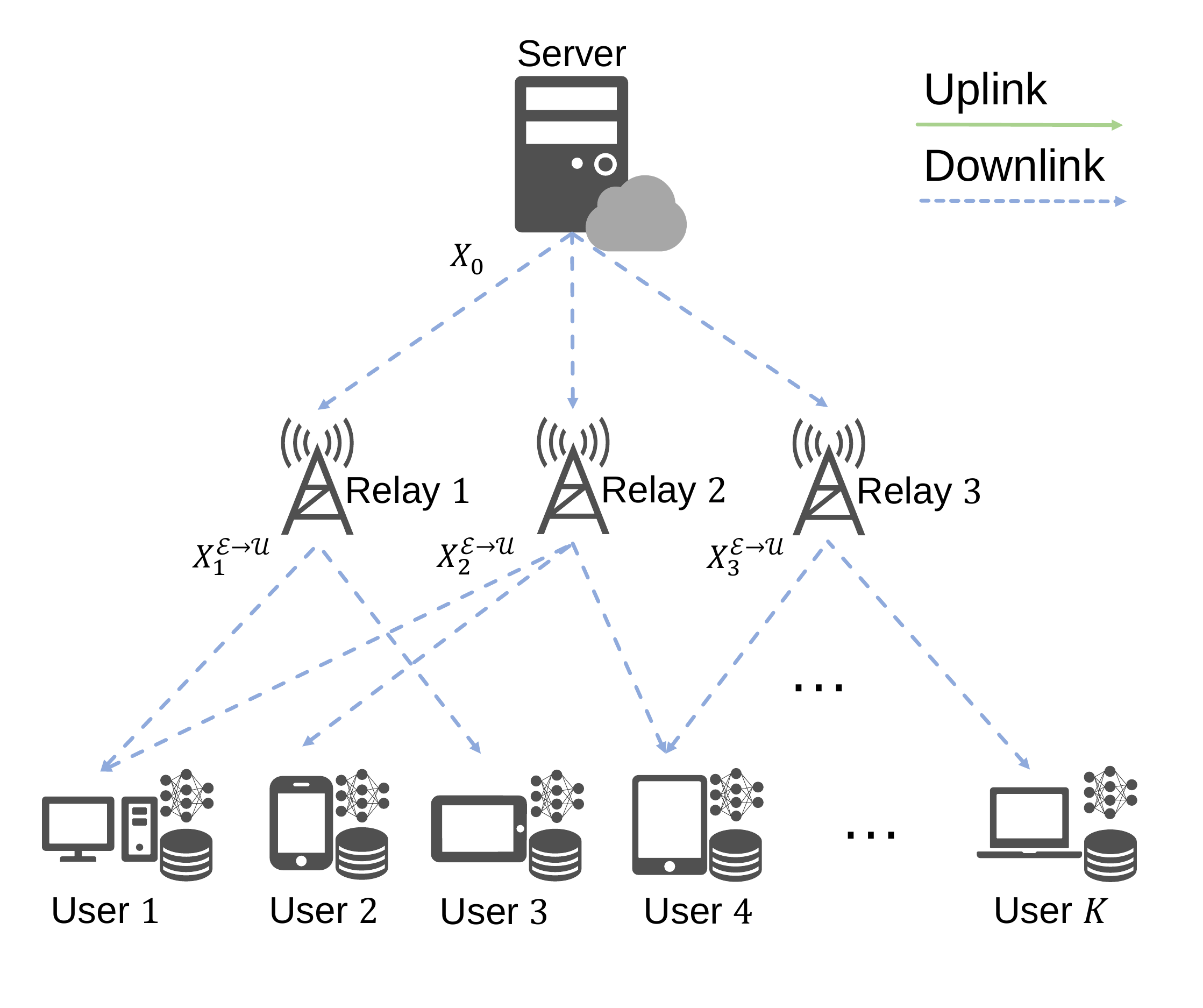}
        }
	\centering
	\caption{A hierarchical distributed MTL system where $K$ users jointly train  multiple models $\{\mathbf{w}_1,\ldots,\mathbf{w}_K\}$ via a layer of $H=3$ relays. At one iteration, each user $k\in[K]$ performs the local update phase and sends the message $X_k^{\mathcal{U}\rightarrow \mathcal{E}}$ to the relays. Based on the received messages, the relay $i\in[3]$ then generates and sends $X_i$ to the server via the uplink. 
    In the downlink, the server sends the message  $X_0$ to the relays, and relay $i\in[3]$ sends  $X_i^{\mathcal{E}\rightarrow \mathcal{U}}$ to the users, such that each user $k$ obtains the updated $\mathbf{w}_k$: (a) Uplink communication; (b)  Downlink communication.}
	\label{fig: SystemModel}
\end{figure}

Additionally, in the practical communication system, the links between remote users and the central server often suffer from limited bandwidth, high latency, and intermittent connections \cite{mao2017survey}.
Direct communication between users and the server can be inefficient, requiring multiple re-transmissions or increased transmission power, which slows down the distributed learning process.
Recently, hierarchical learning frameworks, such as fog computing and mobile edge computing, have been designed to mitigate the problem \cite{liu2020client,prakash2020hierarchical,sasidharan2022coded}, where relay nodes (e.g., pico base stations and edge servers) are added between users and the master server to help to train the model together with the users. The hierarchical learning framework could enlarge the cover range of services, and improve the communication rate between users and the server. Unfortunately, most of the existing works focused on the single-task learning case.  There exists very few works addressing the communication bottleneck problem in hierarchical MTL systems.  How to jointly exploit the hierarchical frameworks and MTL properties to reduce the communication load is still an open problem.

In this paper, we consider the two-hop hierarchical  network architecture, where a layer of $H$ relays connects a central server to $K$ users as depicted in Fig. \ref{fig: SystemModel}.
A network connection matrix $G$ can represent the network connectivity between users and relays, and this network topology can be arbitrarily prefixed.
This two-layer network fits many practical network architectures, such as cellular networks\cite{tse2005fundamentals}, combination networks \cite{ngai2004network,zewail2018combination}, and edge computing systems \cite{liu2020client,prakash2020hierarchical,sasidharan2022coded}. 
The main contributions can be summarized as follows.

\begin{itemize}
    \item We introduce coding techniques to hierarchical MTL systems to mitigate the communication bottleneck problem.  Our proposed coded scheme is feasible for arbitrary network topologies between users and relays. More importantly, unlike previous coded computing techniques which require repetitive storage of data among users that causes the thread of privacy leakage, our coding scheme achieves coded multicast gain by exploiting the network topology, instead of introducing repetitive data placement.  Notice that our scheme is lossless transmission, i.e., at each iteration, the users obtain the same update models as the uncoded scheme without sacrificing the convergence performance.  To the best of our knowledge, this is the first work to use coding techniques to reduce the communication loads for hierarchical MTL systems.
    
    \item Unlike the conventional scheme where relays only forward the received message, in our scheme the relays generate coded symbols according to the network topology, with each symbol intended by many other relays,  thereby obtaining the multicast gain. Also, instead of letting the master server perform the global update to obtain all global models, we let the master server send a linear combination of the coded symbols sent by the relays, by which the relays can first decode their required information and then perform the global update. Finally, the relays send the updated models to the desired users.   From theoretical analysis, we show that our scheme can greatly reduce the communication loads both in the uplink and downlink transmissions between the server and relays. Experiments on real datasets demonstrate that  our proposed scheme can reduce the total training time by 17\% $\sim$ 26\% compared to the conventional uncoded scheme.
    \item We derive the information-theoretic lower bounds on the uplink and downlink communication loads under the hierarchical distributed MTL setting. We show that the gaps between the upper bounds of our proposed scheme and the lower bounds are within  the minimum number of connected users among all relays, which demonstrates the scalability of our scheme. Moreover, under the setting where the network connection matrix can be delicately designed, the proposed scheme can achieve the information-theoretic optimal load pair.
\end{itemize}
%, unlike traditional compression schemes such as sparsification \cite{aji2017sparse,stich2018sparsified,wangni2017gradient}, and quantization \cite{alistarh2017qsgd,bernstein2018signsgd,liang2021improved}

\emph{Related Works:}  The coding technique is a promising approach to alleviate communication bottlenecks while achieving lossless transmission. In the seminal work of coded computation, \cite{li2017fundamental} proposes coded distributed computation (CDC), which can  significantly reduce communication loads by introducing redundant storage and computation to create coded transmission in the communication phase. In   \cite{tang2021communication}, some of the authors have applied the idea of coded transmission to the MTL setting, and   the proposed scheme reduces the communication loads by using redundant placement and computation on the publicly shared dataset to introduce coded multicasting opportunities.
In \cite{prakash2020coded}, structured coding
 is injected into federated learning \cite{kairouz2021advances} for speeding up the training procedure. 
However, the scheme in \cite{tang2021communication} requires a control master to delicately allocate the data across distributed users to enable repetitive storage, which is not applicable when the training data is collected locally by users or when the data is private as in federated learning.
\cite{prakash2020coded} avoids redundant storage but only focuses on the  \emph{single-task}  learning model. Besides, the coded transmission in \cite{prakash2020coded} is lossy, i.e., reducing the communication load at the cost of degrading the learning performance.
Moreover, both \cite{tang2021communication} and \cite{prakash2020coded} consider the single-layer broadcast network, rather than hierarchical frameworks.

The CDC-based methods are closely related to the coded caching strategy \cite{maddah2014fundamental}, as they both use repetitive stored data as side information to create multicast opportunities and reduce communication loads in the network. There exist some works on coded caching considering two-hop networks. The work in \cite{karamchandani2016hierarchical,wang2019reduce} both consider a two-layer network where a central server is connected to $K_1$ mirror sites and each mirror, in turn, is connected to $K_2$ users. Using the memory storage  in mirrors and the users, the communication load of both hops can be reduced via coded transmission. The work of \cite{ji2015fundamental,tang2016coded,zewail2017coded,zewail2019cache,zewail2018combination}  explores coding caching schemes in combination networks \cite{ngai2004network}.
In such networks, the number of users satisfies $K=\binom{H}{r}$ for $r\in[H]$, and each user is connected to a unique subset of relays of size $r$. 
Noting that the topology of the combined network is highly symmetric, it is natural to use the idea of coded caching.
In \cite{ji2015fundamental}, the coded multicasting-combination network coding method is proposed, and  the achievable maximum link load is inversely proportional to the per-user storage capacity and to the degree of each user.
\cite{tang2016coded} considers the coded caching scheme under the setting of combination networks with the resolvability property, i.e., $r$ divides $H$. 
\cite{zewail2017coded} utilizes maximum distance separable (MDS) codes, and achieves the same performance as \cite{tang2016coded} while removing the 
constraint of resolvability.
Moreover, \cite{zewail2019cache} considers the setting with asymmetric end users, and \cite{zewail2018combination} further considers the privacy constraints.
Note that the above methods \cite{karamchandani2016hierarchical,wang2019reduce,ji2015fundamental,tang2016coded,zewail2017coded,zewail2019cache,zewail2018combination} mostly require symmetrical data placement and do not consider how to integrate with MTL.
In addition, unlike \cite{karamchandani2016hierarchical,wang2019reduce,ji2015fundamental,tang2016coded,zewail2017coded,zewail2019cache,zewail2018combination}, in this paper, we consider the topology of the network with arbitrary users and relays, and the above work can all be included. This scenario is highly challenging as we do not consider any symmetric property, which is important in coded caching scheme design. 
%Besides, the optimization-based design we used is also applied to coded schemes in other heterogeneous scenarios \cite{wang2022universal,ibrahim2019coded,daniel2019optimization,wang2019optimization}.

The rest of the paper is organized as follows. Section \ref{sec: system_model} introduces the multi-task learning framework and the system model of the hierarchical system. 
Section \ref{sec: example} uses a motivating  example to show how our scheme reduces communication loads.
Section \ref{sec: scheme} presents the general description of our proposed coded scheme. 
Section \ref{sec: experiement} verifies our scheme through experiments on real-world datasets. Section \ref{sec: conclusion} concludes our paper.% and  some of the proofs are given in the Appendices.

\emph{Notations}: For a positive integer $a$, let $[a]\triangleq \{1,\ldots,a\}$.
Let $\mathbb {R}$ be the set of all real numbers, $\mathbb {N}$ be the set of natural numbers, and $\mathbb {N}^+$ be the set of natural numbers without zero.
$\mathbf{A}^T$ denotes the transpose of matrix $\mathbf{A}$. Mod$(a,b)$ denotes  the modulo operation on $a$ with integer divisor $b$ and in this paper, we let Mod$(a,b)\in[a]$, and particularly Mod$(a,b)=a$ when $b$ divides $a$. $\binom{a}{b}=0$ if $a<0$, $b<0$ or $a<b$. 

\section{System Model and Problem Definition} \label{sec: system_model}
In this section, we first introduce a uniform framework for widely used distributed MTL algorithms, such as CoCoA\cite{jaggi2014communication}, MOCHA\cite{smith2017federated}, FedU\cite{dinh2021new}, and  then introduce the system model.%.

\subsection{Preliminary: Distributed MTL Algorithms}
Consider a distributed  MTL system where  a central server and $K$ distributed users collaboratively train $K\in\mathbb{N}^+$ different tasks.
We denote the dataset at the user $k$ as $\mathcal{B}_k$, consisting of $|\mathcal{B}_k|$ data points with $ \mathbf{x}_{k, j}\in\mathbb{R}^l$ being the $j$-th point and $y_{k,j}$ as its label. Here $ y_{k,j}$  could be continuous for a regression problem or discrete for a classification problem.
Each user wishes to learn a unique  model $ \mathbf{w}_k \in \mathbb{R}^{m}$, for some $m\in\mathbb{N}^+$. %Here $m$ is the dimension of the learning model.

%\footnote{$\ell_{k}(\cdot)$ can be either convex or non-convex. In some optimization algorithms, such as \cite{smith2017federated}, $\ell_{k}(\cdot)$ is restricted  to be convex.}
Consider a general distributed MTL setting introduced in \cite{smith2017federated,jaggi2014communication,dinh2021new}, which can be formulated as the following problem:
\begin{equation}\label{eq1}
\min _{\mathbf{W}, \mathbf{\Omega}}\left\{\sum_{k=1}^{K} \sum_{j=1}^{|\mathcal{B}_{k}|} \ell_{k}\left(\mathbf{w}_{k},  \mathbf{x}_{k,j}, y_{k,j}\right)+\mathcal{R}(\mathbf{W}, \mathbf{\Omega})\right\},
\end{equation}
where $\ell_{k}(\cdot)$ denotes either convex or non-convex loss function of the $k$-th task such as square loss or hinge loss for Suppor Vector Machine (SVM) models,  $\mathbf{W} \triangleq\left[\mathbf{w}_{1}, \mathbf{w}_{2}, \ldots, \mathbf{w}_{K}\right] \in \mathbb{R}^{m \times K}$ is a matrix whose $k$-th column $\mathbf{w}_{k}$ is the model parameters for the $k$-th task, the matrix  $\mathbf{\Omega} \in \mathbb{R}^{K \times K}$ is a correlation matrix modeling relationships among tasks, e.g.,   $\mathbf{\Omega}=(\mathbf{W}^T\mathbf{W})^{\frac{1}{2}}$ in \cite{smith2017federated},  and the  regularization term $\mathcal{R}(\cdot)$ takes  $({\mathbf{W}},\mathbf{\Omega})$ as inputs and  differs in different MTL problems. For example, several popular MTL approaches \cite{zhang2010convex,zhou2011clustered,evgeniou2004regularized,jacob2008clustered} use the  bi-convex formulation
$\label{Regularization}
\mathcal{R}(\mathbf{W}, {\mathbf{\Omega}}) = \lambda_1\text{tr} ({\mathbf{W}}{\mathbf{\Omega}} {\mathbf{W}}^T)+\lambda_2\|\mathbf{W}\|^2_{F},
$
for some constants $\lambda_1, \lambda_2\ge 0$, and $\| \cdot \|_F$ denotes the Frobenius norm. Note that for the MTL framework where the correlation between local models is not considered, we can set $\lambda_1=0$. The training of distributed MTL  contains two phases: local update and global update.  %For simplicity, we focus on one iteration and omit the iteration index.
%To solve the problem (\ref{eq1}), several algorithms are proposed  including CoCoA\cite{jaggi2014communication}, MOCHA\cite{smith2017federated}, and FedU\cite{dinh2021new}.  %For instance,   \cite{smith2017federated} proposed an alternating optimization procedure that consists two steps: solving $\mathbf{W}$ with a fixed $\mathbf{\Omega}$ in the distributed manner, and solving $\mathbf{\Omega}$ with aggregated $\mathbf{W}$ from all users on the server. %The distributed multi-task learning optimization method is given in Algorithm \ref{alg_optimization}.

\textbf{Local Update:} 
At each iteration, user $k \in [K]$ first executes local training to generate IVs based on the local data $\mathcal{B}_k$ and the model parameters $\{\mathbf{w}_k\}$ of the previous iteration, i.e., user $k$ computes
\begin{align}\label{equ:local_update}
    \mathbf{v}_k={f}_k(\mathcal{B}_k,\mathbf{w}_k),
\end{align}
where the local update function ${f_k}:\mathbb{R}^{l\times|\mathcal{B}_k|} \times \mathbb{F}_{2^V} \rightarrow \mathbb{F}_{2^V}$ maps $(\mathcal{B}_k,\mathbf{w}_k)$ into the IV  $\mathbf{v}_k\in \mathbb{F}_{2^V}$, 
where $V\in \mathbb{N}^+$ denotes the size of $ \mathbf{v}_k$ and $\mathbf{w}_k$ after the quantization process\footnote{The local update $ \mathbf{v}_k$ is a continuous value and should be compressed before transmission. There exists comprehensive research on lossy compression, which is beyond  the focus of this paper.}, i.e., $\mathbf{w}_k \in \mathbb{F}_{2^V}$.  For example,   \cite{smith2017federated} considered a distributed primal-dual optimization of the problem \eqref{eq1}, and each user $k$ independently solves a subproblem  to  obtain the IV $\mathbf{v}_{k}$. 

\textbf{Global Update}:
The node that is responsible for performing the global update (e.g., the server)  first recovers all IVs  $\{\mathbf{v}_1,\ldots,\mathbf{v}_{K}\}$, and then updates the global models   $(\mathbf{w}_1,\ldots,\mathbf{w}_{K})$ as follows
%Then  the node  performs the global update based on all IVs, i.e.,
\begin{align} \label{equ:global_update}
    (\mathbf{w}_1,\ldots,\mathbf{w}_{K})=\phi(\mathbf{v}_1,\cdots,\mathbf{v}_{K}),
\end{align}
where $\phi: (\mathbb{F}_{2^{V}})^K\rightarrow (\mathbb{F}_{2^{V}})^K$ is the global update function. For example,  the global update function in  \cite{smith2017federated} is defined as  
%\begin{subequations}\label{ReduceOpera}
%\begin{IEEEeqnarray}{rCl}\label{ReduceOpera1}
$ (\mathbf{w}_1,\ldots,\mathbf{w}_{K})=\nabla \mathcal{R}^{*}([\mathbf{v}_1,\ldots, \mathbf{v}_K ]).$ 
%\end{IEEEeqnarray} 

%considered a distributed primal-dual optimization of the problem \eqref{eq1}, and each user $k$ independently solves a subproblem  to  obtain the IV $\mathbf{v}_{k}$. 

%In the local update, every user $k$   solves a local dual subproblem with fixed $\mathbf{\Omega}$ and obtains  an \textit{intermediate value}  $\mathbf{v}_{k}\in(\mathbb{F}_{2^F})^{d}$, which can be formalized  as 
%\begin{IEEEeqnarray}{rCl}
%\mathbf{v}_{k}=f_k(\mathcal{B}_{k}, \mathbf{\Omega}),
%\end{IEEEeqnarray}
%where $f_k$ is a local update function at user $k$.

%In the global update phase, the  ensemble of intermediate values  $\{\mathbf{v}_{k}\}_{k=1}^K$  will be used to update $\mathbf{W}$ and $\mathbf{\Omega}$, which can be formalized as  
%\begin{IEEEeqnarray}{rCl}\label{eqReduce}
%(\mathbf{W},\mathbf{\Omega})=\phi(\mathbf{v}_{1},\ldots,\mathbf{v}_{K} ),
%\end{IEEEeqnarray}
%where   $\phi$ is a global update function. 

\subsection{System  Model}

In this subsection, we introduce the communication model of the hierarchical  MTL framework, in which $K$ users compute a single output function from $K $ sets of input data $\{\mathcal{B}_{1}, \ldots \mathcal{B}_{K}\}$ with the help of a server and $H\in \mathbb{N}^+$ relays. We assume that the relays are equipped with some computational ability as in \cite{liu2020client,prakash2020hierarchical,sasidharan2022coded}.

\subsubsection{Network Topology }
We consider a two-hop network, where the server $S$, is connected to $K$ end users via a set of $H$ relays. 
We denote the set of users as $\mathcal{U}$, and the set of relays as $\mathcal{E}$.
 As illustrated in Fig. \ref{fig: SystemModel}, all relays are connected to the server through an error-free shared link. 
We define a network connection matrix $G\in \mathbb{N}^{H\times K}$ to show the connection between relays and users \cite{biggs1993algebraic}. More specially, $G(i,k)=1$ if the relay $i$ is connected with the user $k$, otherwise $G(i,k)=0$, for all $i\in[H]$ and $ k\in [K]$.
We assume all available network links between relays and users are assumed to be noiseless.
We denote the indices of users connected to the relay $i\in [H]$ as $\mathcal{N}_i$, where $\mathcal{N}_i \subseteq [K]$. 
The connection topology between the relays and the users  is arbitrary but fixed, that is, it can not be designed. 
To ensure that all models are trained based on all training datasets $\{\mathcal{B}_{1}, \ldots \mathcal{B}_{K}\}$, we assume that  $\cup_{i\in [H]}\mathcal{N}_i=[K].$ Also, if there exists a relay connecting all users, then there is no necessary to use the hierarchical network as this relay can serve as a master server. Thus, we consider the nontrivial case where $\mathcal{N}_i\neq [K]$ for all $i\in[H]$.

To further characterize the network connection, we  define the user connectivity, the average user connectivity, and $z$-connected users as follows. 
\begin{Definition}[User Connectivity and Average User Connectivity]
The user connectivity, denoted as $r_k$, for $k\in[K]$, indicates the number of relays that user $k$ connects. The average user connectivity denoted as $r$, indicates the total number of links between relays and users, normalized by the number of users $K$, i.e., $r=\frac{1}{K}\sum_{k=1}^K r_k.$

\end{Definition}

\begin{Definition}[$z$-Connected Users]
The $z$-connected users, denoted by $\mathcal{I}_z$, indicates the indices of users \emph{exclusively} connected by $z\in[H]$ relays, i.e., 
\begin{IEEEeqnarray}{rcl}\label{DefIz}
	\mathcal{I}_z\triangleq\left\{j:\mathcal{C} \subseteq [H],|\mathcal{C}|=z,j\in\underset{i\in\mathcal{C}}{\cap}\mathcal{N}_i, j\notin \underset{i\in [H] /\ \mathcal{C}}{\cup}\mathcal{N}_i\right\}.
\end{IEEEeqnarray}
\end{Definition} 

%The learning model and communication model for the hierarchical MTL framework are described as follows. 
%The distributed MTL framework can be decomposed into three phases: the local update phase, the global update phase and the communication phase. 

%\subsubsection{Learning Model}The learning model contains two phases: local update and global update. 

\subsubsection{Learning Model}\label{sec:learngmodel}
 
We consider the general MTL model with arbitrary update and global update functions of forms  \eqref{equ:local_update} and \eqref{equ:global_update}, respectively. This includes the popular MTL algorithms including  CoCoA\cite{jaggi2014communication}, MOCHA\cite{smith2017federated}, FedU\cite{dinh2021new}, etc., which are proposed for single-layer networks without relay nodes, and in which the server performs the global update. For our considered  hierarchical  MTL framework, to ensure the updated models available at all users in any network topology, every relay must obtain all global updated models $(\mathbf{w}_1,\ldots,\mathbf{w}_{K})$\footnote{We may allow users to perform the global update but this will push the users to consume more energy, computation, and storage resources, and thus is not considered.}. 
Therefore, the global update should be performed on relays or the server.  

Furthermore, it is easy to show that performing the global update at relays could incur less communication cost than at the server. This is because the server does not know any  of $\{\mathbf{v}_1,\ldots,\mathbf{v}_{K}\}$ and the relays do not know any  of $ \{\mathbf{w}_1,\ldots,\mathbf{w}_{K}\}$,  if the global update is performed at the server, $KV$ bits are required for both uplink and downlink communications between relays and the server. %Thus in the subsequent sections, we only focus on reducing the communication load pair between relay nodes and the server, i.e., $(L_\textnormal{up}^{\mathcal{E}\rightarrow S}, L_\textnormal{down}^{S\rightarrow \mathcal{E}})$.  
To reduce the communication cost, we assume relays perform the global update This is a reasonable assumption, as the relays are often equipped with  computational ability in many settings \cite{liu2020client,prakash2020hierarchical,sasidharan2022coded}.

\subsubsection{Communication Model}
The communication phase consists of two stages: uplink and downlink communications. The uplink communication consists of the communication from users to relays and from relays to the server. 
The downlink communication consists of the communication from the server to relays and  from relays to users. 
The goal of the communication  is ensure the each user $k$ obtains desired model $\mathbf{w}_k$, for all $k\in[K]$ .

\textbf{Uplink Communication:} Based on the local update $\mathbf{v}_{k}$, user $k \in [K]$  generates and sends  message  $X_k^{\mathcal{U}\rightarrow \mathcal{E}} \in \mathbb{F}_{2^{T_{\mathcal{U}_k}}},$ for some $T_{\mathcal{U}_k} \in \mathbb{N}^+$,  to the set of relays $\{i:k \in \mathcal{N}_i\}$ through the uplink, i.e.,
\begin{equation}
    X_k^{\mathcal{U}\rightarrow \mathcal{E}} = \psi^\mathcal{U}_{k}\left(\mathbf{v}_{k}\right), 
\end{equation}
where $\psi^\mathcal{U}_{k}:\mathbb{F}_{2^V}\rightarrow \mathbb{F}_{2^{T_{\mathcal{U}_k}}}$ is the encoding function at user $k$ for the communication to the corresponding relays. 

Denote  $\mathcal{V}_{i}$ as the  set of local IVs obtained by relay $i\in[H]$ based on the received messages from the users, and we have
\begin{IEEEeqnarray}{rcl}
    \mathcal{V}_{i}=\phi_{i}^\mathcal{E}\left(\{X_k^{\mathcal{U}\rightarrow \mathcal{E}}: k\in \mathcal{N}_i\} \right),
\end{IEEEeqnarray}
where $\phi_{i}^\mathcal{E}: \prod_{k\in \mathcal{N}_i} \mathbb{F}_{2^{T_{\mathcal{U}_k}}}\rightarrow (\mathbb{F}_{2^{V}})^{|\mathcal{V}_{i}|}$ is the decoding function at relay $i$.

Based on  received messages from user $k\in \mathcal{N}_i$, relay $i\in[H]$ generates and sends  message  $X_i \in \mathbb{F}_{2^{T_{i}}},$ for some $T_{i} \in \mathbb{N}^+$,  to the server through the shared uplink, i.e.,
\begin{equation}
    X_i = \psi_{i}\left(\{X_k^{\mathcal{U}\rightarrow \mathcal{E}}: k\in \mathcal{N}_i\}\right), 
\end{equation}
where $\psi_{i}: \prod_{k\in \mathcal{N}_i} \mathbb{F}_{2^{T_{\mathcal{U}_k}}}\rightarrow \mathbb{F}_{2^{T_i}}$ is the encoding function at relay $i$ in the uplink. 

\begin{Definition}[Uplink Communication Loads]
  We define the uplink communication load from users to relays and from relays to the server, denoted by $L_\textnormal{up}^{\mathcal{U}\rightarrow \mathcal{E}}$ and $ L_\textnormal{up}^{\mathcal{E}\rightarrow S}$ respectively, as the total number of bits sent in the uplink at each iteration, normalized by the size of a single  IV, i.e., $  L_\textnormal{up}^{\mathcal{U}\rightarrow \mathcal{E}}  \triangleq \frac{\sum_{k=1}^K T_{\mathcal{U}_k}}{V}$ and
          $L_\textnormal{up}^{\mathcal{E}\rightarrow S}  \triangleq \frac{\sum_{i=1}^H T_i}{V}.$
%\begin{IEEEeqnarray}{rCl}
 % L_\textnormal{up}^{\mathcal{U}\rightarrow \mathcal{E}}  \triangleq \frac{\sum_{k=1}^K T_{\mathcal{U}_k}}{V}, ~
  %        L_\textnormal{up}^{\mathcal{E}\rightarrow S}  \triangleq \frac{\sum_{i=1}^H T_i}{V}.
%\end{IEEEeqnarray}

\end{Definition}

%Notice that when the central server performs the global update, it needs to recover all IVs using the uplink messages from the relays, i.e.,
%\begin{align}
%    (\mathbf{v}_1,\ldots,\mathbf{v}_{K})=\varphi_{0}\left(X_1, X_2, \ldots, X_H\right),
%\end{align} 
%where  $\varphi_{0}:\mathbb{F}_{2^{T_1}}\times\cdots \times \mathbb{F}_{2^{T_H}}\rightarrow  (\mathbb{F}_{2^{V}})^K$ is the decoding function at the server.

\textbf{Downlink Communication:} 
Based on the  received   uplink messages $(X_1, X_2, \ldots, X_H)$, the server generates message $X_{0}\in \mathbb{F}_{2^{T_0}},$ for some $T_0 \in \mathbb{N}^+$, and  broadcasts it to all relays through the  downlink at each iteration,
\begin{equation}
    X_{0} = \psi_0(X_1, X_2, \ldots, X_H),
\end{equation}
where $\psi_0: \mathbb{F}_{2^{T_1}}\times\cdots \times \mathbb{F}_{2^{T_H}}\rightarrow \mathbb{F}_{2^{T_0}} $ is the encoding function at the central server. 

%$i\in [H]$ 
As mentioned in Section \ref{sec:learngmodel}, all the relays perform the global update, and thus  each relay $i\in [H]$  recovers all IVs using the  message $X_{0}$ sent from the server and the local IVs $\mathcal{V}_i$, i.e.,
\begin{align}
    (\mathbf{v}_1,\ldots,\mathbf{v}_{K})=\varphi_{i}\left(X_{0}, \mathcal{V}_i\right),
\end{align} 
where  $\varphi_{i}:\mathbb{F}_{2^{T_0}} \times (\mathbb{F}_{2^V})^{|\mathcal{V}_i|} \rightarrow  (\mathbb{F}_{2^{V}})^K$ is the decoding function at   relay $i$.

Next, based on the received downlink message $X_{0}$ and local IVs $\mathcal{V}_i$, each relay $i\in[H]$ generates and sends message  $X_{i}^{\mathcal{E} \rightarrow \mathcal{U}} \in \mathbb{F}_{2^{T_{\mathcal{E}_{i}}}},$ for some $T_{\mathcal{E}_{i}} \in \mathbb{N}^+$,  to the set of users $\{k: k \in \mathcal{N}_i\}$ through the downlink, i.e.,
\begin{equation}
    X_{i}^{\mathcal{E} \rightarrow \mathcal{U}}  = \psi^\mathcal{E}_i \left(X_{0}, \mathcal{V}_{i}\right), 
\end{equation}
where $\psi^\mathcal{E}_i:\mathbb{F}_{2^{T_0}} \times (\mathbb{F}_{2^{V}})^{|\mathcal{V}_{i}|} \rightarrow \mathbb{F}_{2^{T_{\mathcal{E}_{i}}}}$ is the encoding function at relay $i$ for the communication to the corresponding users. 

Finally, each user $k\in[K]$ recovers the desired learning model $\mathbf{w}_k$ according to the received messages from the connected relays, i.e.,
\begin{align}
    \mathbf{w}_k = \phi_{k}^\mathcal{U} (\{X_{i}^{\mathcal{E} \rightarrow \mathcal{U}}: k \in \mathcal{N}_i\}),
\end{align}
where $\phi_{k}^\mathcal{U}: \prod_{k\in \mathcal{N}_i} \mathbb{F}_{2^{T_{\mathcal{E}_{i}}}}\rightarrow \mathbb{F}_{2^{V}}$ is the decoding function at user $k$. 
 
\begin{Definition}[Downlink Communication Loads] 
We define the downlink communication load from the server to relays and from relays to users, denoted by $L_\textnormal{down}^{S\rightarrow \mathcal{E}}$ and $L_\textnormal{down}^{\mathcal{E}\rightarrow \mathcal{U}}$ respectively, as the  number of bits sent in the downlink at each iteration, normalized by the size of a single  IV, i.e., $ L_\textnormal{down}^{S\rightarrow \mathcal{E}} \triangleq\frac{T_0}{V}$,  and 
        $L_\textnormal{down}^{\mathcal{E}\rightarrow \mathcal{U}}  \triangleq \frac{\sum_{i=1}^H  T_{\mathcal{E}_{i}}}{V}.$
%\begin{IEEEeqnarray}{rCl}
  %      L_\textnormal{down}^{S\rightarrow \mathcal{E}} \triangleq\frac{T_0}{V}, ~
  %      L_\textnormal{down}^{\mathcal{E}\rightarrow \mathcal{U}}  \triangleq \frac{\sum_{i=1}^H  T_{\mathcal{E}_{i}}}{V}.
%\end{IEEEeqnarray}
\end{Definition}

The  communication loads  $(L_\textnormal{up}^{\mathcal{U}\rightarrow \mathcal{E}}, L_\textnormal{up}^{\mathcal{E}\rightarrow S}, L_\textnormal{down}^{S\rightarrow \mathcal{E}}, L_\textnormal{down}^{\mathcal{E}\rightarrow \mathcal{U}})$ are said to be \emph{achievable} if there exists a scheme consisting of encoders $\{\psi_{k}^\mathcal{U}\}_{k=1}^K, \{\psi_{i}\}_{i=0}^H, \{\psi_{i}^\mathcal{E}\}_{i=1}^H$ and decoders $\{\phi_k^\mathcal{U}\}_{k=1}^K$, $\{\phi_i^\mathcal{E}\}_{i=1}^H, \{\varphi_{i}\}_{i=0}^H$ such that each user $k \in[K]$ can successfully obtain the desired $\mathbf{w}_{k}$. In this paper, our goal is to reduce the (achievable) communication loads for any given topological connection, and even find the optimal scheme for minimizing the communication loads for some cases.

\begin{Remark}
    To obtain the desired $(\mathbf{w}_1,\ldots,\mathbf{w}_{K})$, each user $k \in[K]$ needs to send the IV $\mathbf{v}_k$ and download the update model $\mathbf{w}_k$ via its connected relays. Because the relays do not perform the local update and there is no repetitive data placement among users, the IVs and models need to be unicasted via the uplink and downlink, resulting in  communication load between users and relays $L_\textnormal{up}^{\mathcal{U}\rightarrow \mathcal{E}}=L_\textnormal{down}^{\mathcal{E}\rightarrow \mathcal{U}}=K$. Thus we only focus on reducing the communication load pair between relays and the server, i.e., $(L_\textnormal{up}^{\mathcal{E}\rightarrow S}, L_\textnormal{down}^{S\rightarrow \mathcal{E}})$.
\end{Remark}

% Since in the uplink communication, the server does not know any value of $ \{\mathbf{v}_1,\ldots,\mathbf{v}_{K}\}$ and the relays do not know any value of $ \{\mathbf{W}_1,\ldots,\mathbf{W}_{K}\}$,  if the global update is performed at the server, it's easy to obtain the downlink and uplink communication loads   $(L_\textnormal{up}^{\mathcal{E}\rightarrow S}, L_\textnormal{down}^{S\rightarrow \mathcal{E}})=(K,K)$. %Thus in the subsequent sections, we only focus on reducing the communication load pair between relays and the server, i.e., $(L_\textnormal{up}^{\mathcal{E}\rightarrow S}, L_\textnormal{down}^{S\rightarrow \mathcal{E}})$.  
%  To reduce the communication loads, we allow the relay to perform the global update\footnote{It can be shown that if we allow the user to perform the global update, we can slightly further reduce the communication loads. However, this will push the users to consume more energy, computing, and storage resource, and thus is not considered.}. 

Denote the  execution time of a single iteration as $T_{\text{total}}$, which roughly consists of computation time $T_{\text{comp}}$ and communication time $T_{\text{comm}}$, i.e.,
\begin{align}\label{equ: total}
 T_{\text{total}} = T_{\text{comp}} + T_{\text{comm}},
\end{align}
where $T_{\text{comp}}$ denotes the time spent on updating parameters (local and global updates) and the coding overheads at each iteration, and $T_{\text{comm}}$ denotes the communication time at each iteration. 
The communication time is roughly estimated as the number of bits sent divided by the bandwidth

We assume that the communication bandwidth between users and relays and between relays and the server is $W_1$ bps and $W_2$ bps respectively.
The communication time can be roughly calculated as the total number of transmitted bits  divided by the bandwidth $W_1$ and $W_2$ \cite{smith2017federated,huang2013depth}, i.e., 
\begin{align} \label{equ: comm}
 T_{\text{comm}} = \frac{(L_\textnormal{up}^{\mathcal{U}\rightarrow \mathcal{E}}+ L_\textnormal{down}^{\mathcal{E}\rightarrow \mathcal{U}})\cdot V}{W_1} + \frac{(L_\textnormal{up}^{\mathcal{E}\rightarrow S}+ L_\textnormal{down}^{S\rightarrow \mathcal{E}})\cdot V}{W_2}.
\end{align}
where $(L_\textnormal{up}^{\mathcal{U}\rightarrow \mathcal{E}}, L_\textnormal{up}^{\mathcal{E}\rightarrow S}, L_\textnormal{down}^{S\rightarrow \mathcal{E}}, L_\textnormal{down}^{\mathcal{E}\rightarrow \mathcal{U}})$ are the achievable communication loads, and $V$ is the bit size of each IV. 

\begin{Example}[Uncoded Scheme (MOCHA\cite{smith2017federated})] \label{example: uncoded}
For example, in the original distributed multi-task learning algorithm, each user $k \in[K]$ needs to send $\mathbf{v}_k$ to the relays, and then the relays directly forward the received messages to the central server. Thus the total number of bits sent by users and relays is $K ·V$ in the uplink, leading to the overall uplink communication load $L_\textnormal{up}^{\mathcal{U}\rightarrow \mathcal{E}} = L_\textnormal{up}^{\mathcal{E}\rightarrow S} = K \cdot V /V=K.$ 

The global update   in \cite{smith2017federated} is only executed on the server. 
After recovering all the IVs, the server computes the global update function as 
$
(\mathbf{w}_1,\ldots,\mathbf{w}_{K})=\phi(\mathbf{v}_1,\cdots,\mathbf{v}_{K}),
$
e.g., the global update function $\phi$ contains two steps as shown in \cite{smith2017federated}.
In the downlink, the central  server sends $\{\mathbf{w}_k\}_{k=1}^K$  to relays, and then relays forwards the received message to the corresponding user.
Thus the total number of bits sent by the server and relays is $K\cdot V$, and the downlink communication load $L_\textnormal{down}^{S\rightarrow \mathcal{E}}= L_\textnormal{down}^{\mathcal{E}\rightarrow \mathcal{U}}=K\cdot V/V =K$.
We refer to this hierarchical distributed MTL method as the uncoded scheme, and obtain its achievable communication loads as:
\begin{align}\label{eqUncoded}
    ( L_\textnormal{up,uncoded}^{\mathcal{E}\rightarrow S},L_\textnormal{down,uncoded}^{S\rightarrow \mathcal{E}})= (K,K).
\end{align}

The time costs at each iteration, denoted by $T_{\text{total}}^{\text{uncoded}}$, is
\begin{align} 
    T_{\text{total}}^{\text{uncoded}} = T_{\text{comp}}^{\text{uncoded}} + \frac{2KV}{W_1}+\frac{2KV}{W_2},
\end{align}
where  $T_{\text{comp}}^{\text{uncoded}}$ denotes the time MOCHA spends for computation at each iteration.
\end{Example}

\section{A Motivating Example}\label{sec: example}
%In this section, we first introduce a motivating example to show how to reduce communication loads by  coded transmission.the network connection  matrix $G$ is given as follows,
%\begin{IEEEeqnarray}{RCL}
%    {G}=
%\begin{bmatrix}
%1 & 0 & 1 & 0 & 1 & 1 & 0 & 1 & 1 & 1\\
%1 & 1 & 1 & 1 & 0 & 0 & 0 & 0 & 0 & 1\\
%0 & 1 & 0 & 1 & 1 & 0 & 0 & 0 & 0 & 0\\
%0 & 0 & 0 & 0 & 0 & 1 & 1 & 0 & 1 & 1\\
%1 & 1 & 0 & 0 & 0 & 1 & 1 & 1 & 0 & 0
%\end{bmatrix},
%\end{IEEEeqnarray}
%which means the corresponding network connection indices 
\begin{figure}[htbp]
	\centering
	\includegraphics[scale=0.7]{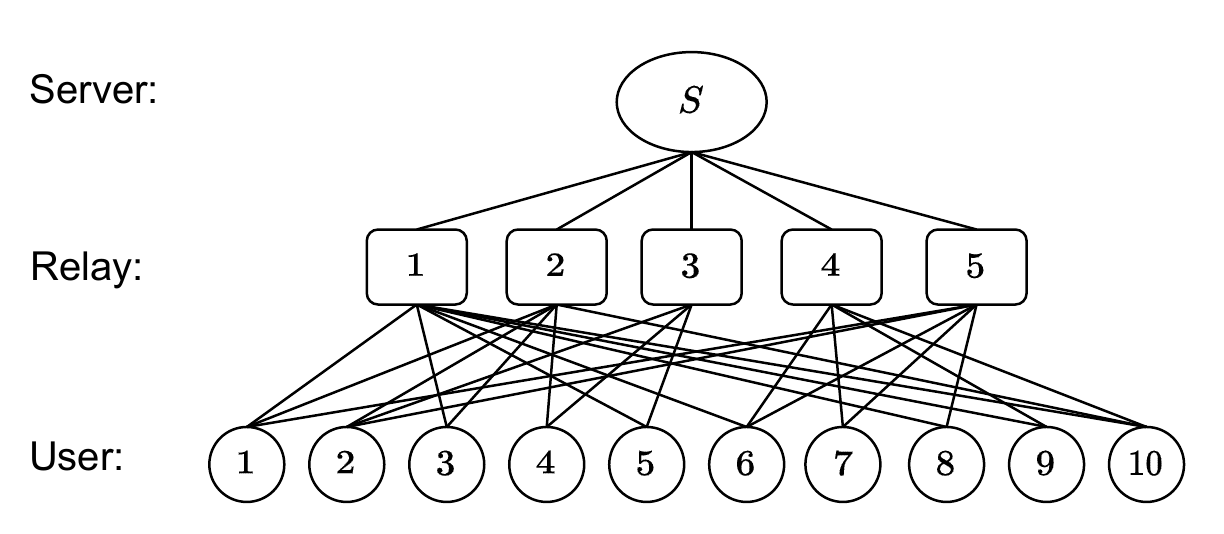}
	\caption{A hierarchical network with $K=10$ users, $H=5$ relays, and network connection matrix $G$. }
	\label{fig: hcmtl_example}
\end{figure}

Consider a hierarchical distributed MTL system where $K=10$ users wish
to learn separate models via a layer of $H=5$ relays,  as illustrated in Fig. \ref{fig: hcmtl_example}. In this example, we have 
$\mathcal{N}_1=\{1,3,5,6,8,9,10\}$, 
$\mathcal{N}_2=\{1,2,3,4,10\}$, 
$\mathcal{N}_3=\{2,4,5\}$,  
$\mathcal{N}_4=\{6,7,9,10\}$, 
$\mathcal{N}_5=\{1,2,6,7,8\}$, and the   $z$-connected users  
$\mathcal{I}_2=\{3,4,5,7,8,9\}$, 
$\mathcal{I}_3=\{1,2,6,10\}$, and $\mathcal{I}_1=\mathcal{I}_4=\emptyset$.

\textbf{Uplink Communication:}
During the local training phase, user $k\in[K]$ generates the IV $\mathbf{v}_k$, and then transmits $\mathbf{v}_k$ to the set of relays $\{i:k \in \mathcal{N}_i\}$ via the uplink, i.e., $X_k^{\mathcal{U}\rightarrow \mathcal{E}}=\mathbf{v}_k$. For instance, user 1 sends $\mathbf{v}_1$ to the connected relays, i.e., $X_1^{\mathcal{U}\rightarrow \mathcal{E}}=\mathbf{v}_1$.
Hence the relays obtain the set of IVs, 
$\mathcal{V}_{1}=\{\mathbf{v}_1, \mathbf{v}_3, \mathbf{v}_5, \mathbf{v}_6, \mathbf{v}_8\, \mathbf{v}_9, \mathbf{v}_{10}\}$, 
$\mathcal{V}_{2}=\{\mathbf{v}_1, \mathbf{v}_2, \mathbf{v}_3, \mathbf{v}_4, \mathbf{v}_{10}\}$, 
$\mathcal{V}_{3}=\{\mathbf{v}_2, \mathbf{v}_4, \mathbf{v}_5\}$, 
$\mathcal{V}_{4}=\{\mathbf{v}_6, \mathbf{v}_7, \mathbf{v}_9, \mathbf{v}_{10}\}$ 
and $\mathcal{V}_{5}=\{\mathbf{v}_1, \mathbf{v}_2, \mathbf{v}_6, \mathbf{v}_7, \mathbf{v}_8\}$. 

We decompose the communication phase between relays and the server into multiple rounds, indexed as $z\in [H-1]$, and the $z$ communication round   aims to send IVs $\{\mathbf{v}_k: k\in \mathcal{I}_z\}$.

With respect to IVs $\{\mathbf{v}_k: k\in \mathcal{I}_z\}$, we define $n_i^z$ as the maximum number of IVs available at relay $i$ but unavailable at some other relay $j \in [H]\backslash\{i\}$, i.e.,
\begin{IEEEeqnarray}{rcl} \label{equ: unknown}
    n_i^z=|\mathcal{N}_i \cap \mathcal{I}_z | - \min_{j \in [H]}|\mathcal{N}_i\cap \mathcal{N}_j \cap \mathcal{I}_z|. 
\end{IEEEeqnarray}

%We decompose each  IV in $\mathcal{I}_2$ into 2 segments, and the index of the segment indicates which relay is responsible to send this segment. For example, $\mathbf{v}_3=(\mathbf{v}_{3,1}, \mathbf{v}_{3,2})$, where $\mathbf{v}_{3,1}$ is the relay 1 responsible for sending and $\mathbf{v}_{3,2}$ is the relay 2 responsible for sending. As we defined before, the number of bits in $\mathbf{v}_{3,1}$ is $\alpha^2_1 V$, hence we need $\{\alpha^2_i\}_{i=1}^H$ to guide the decomposition of IV.
For the $z=2$ round, we have $n_1^2=3$, $n_2^2=2$, $n_3^2=2$, $n_4^2=2$ and $n_5^2=2$ according to  (\ref{equ: unknown}).
Solving the optimization problem  in  (\ref{equ: optimization_problem}), we have $\alpha^2_1=\alpha^2_2=\alpha^2_3=\alpha^2_4=\alpha^2_5=\frac{1}{2}$, where $\alpha^2_i$ indicates that the relay $i$ will send  $\alpha^2_iV$ bits of symbols. Based on the optimal $\{\alpha^2_i\}_{i=1}^H$, we divide each $\mathbf{v}_k$ with $k \in \mathcal{I}_2=\{3,4,5,7,8,9\}$ into 2 disjoint equal segments as follows
\begin{IEEEeqnarray}{rcl}
    &&\mathbf{v}_3=(\mathbf{v}_{3,1}, \mathbf{v}_{3,2}),
    \mathbf{v}_4=(\mathbf{v}_{4,2}, \mathbf{v}_{4,3}),
    \mathbf{v}_5=(\mathbf{v}_{5,1}, \mathbf{v}_{5,3}),\nonumber \\ 
    &&\mathbf{v}_7=(\mathbf{v}_{7,4}, \mathbf{v}_{7,5}),
    \mathbf{v}_8=(\mathbf{v}_{8,1}, \mathbf{v}_{8,5}),
    \mathbf{v}_9=(\mathbf{v}_{9,1}, \mathbf{v}_{9,4}),
\end{IEEEeqnarray}
where  the number of bits in $\mathbf{v}_{k,i}$ is $\alpha^2_i V$ for all $k\in \mathcal{I}_2$, and $\mathbf{v}_{k,i}$ indicates that this segment is generated from $\mathbf{v}_{k}$ and will be sent by relay $i$. Under this setting, we divide the segments equally. 
Each relay $i\in[5]$ generates and transmits $n_i^2$ random linear combinations of the segments $\{\mathbf{v}_{k,i}: k \in \mathcal{N}_i\cap \mathcal{I}_2\}$. For instance, relay $1$ sends:
\begin{IEEEeqnarray}{rcl}
    X_{1,j}^{2} = C_{1,j}^{2}(\mathbf{v}_{3,1}, \mathbf{v}_{8,1}, \mathbf{v}_{9,1}),
\end{IEEEeqnarray}
for $j=1,2,3$, where $\{C_{1,j}^{2}(\cdot)\}_{j=1}^3$ are random linear combining functions.

For the $z=3$ round, we have $n_1^3=3$, $n_2^3=2$, $n_3^3=1$, $n_4^3=2$ and $n_5^3=2$ according to (\ref{equ: unknown}). 
Solving the optimization problem in (\ref{equ: optimization_problem}), we have $\alpha^3_1=\alpha^3_3=0$ and $ \alpha^3_2=\alpha^3_4=\alpha^3_5=\frac{1}{2}$. 
We divide each $\mathbf{v}_k$ with $k \in \mathcal{I}_3=\{1,2,6,10\}$ into 3 disjoint equal segments, i.e., 
\begin{IEEEeqnarray}{rcl}
    &&\mathbf{v}_1=(\mathbf{v}_{1,1}, \mathbf{v}_{1,2}, \mathbf{v}_{1,5}), \mathbf{v}_2=(\mathbf{v}_{2,2}, \mathbf{v}_{2,3}, \mathbf{v}_{2,5}),\nonumber\\ 
    &&\mathbf{v}_6=(\mathbf{v}_{6,1}, \mathbf{v}_{6,4}, \mathbf{v}_{6,5}),
    \mathbf{v}_{10}=(\mathbf{v}_{10,1}, \mathbf{v}_{10,2}, \mathbf{v}_{10,4}),
\end{IEEEeqnarray}
where the number of bits in $\mathbf{v}_{k,i}$ is $\alpha^3_i V$ for all $k\in \mathcal{I}_3$. Since $\alpha^3_1=\alpha^3_3=0$, each IV above is in fact  divided into $2$ segments, i.e., $\mathbf{v}_{1,1}=\mathbf{v}_{2,3}=\mathbf{v}_{6,1}=\mathbf{v}_{10,1}=\emptyset$, and  relay $1$ and relay $3$ send nothing in the $z=3$ round.
 
Each relay $i\in \{2,4,5\}$ generates and transmits $n_i^3$ random linear combinations of the segments $\{\mathbf{v}_{k,i}: k \in \mathcal{N}_i\cap \mathcal{I}_3\}$. For instance, relay $2$ sends:
\begin{IEEEeqnarray}{rcl}
    X_{2,j}^{3} = C_{2,j}^{3}(\mathbf{v}_{1,2}, \mathbf{v}_{2,2}, \mathbf{v}_{10,2}),
\end{IEEEeqnarray}
for $j=1,2$, where $\{C_{2,j}^{3}(\cdot)\}_{j=1}^2$ are random linear combining functions.

Hence, using this coding technique, we have the uplink communication load $L_{\textnormal{up}}^{\mathcal{E}\rightarrow S}= (\frac{1}{2}\times 3+ \frac{1}{2}\times 2+\frac{1}{2}\times 2+\frac{1}{2}\times 2+\frac{1}{2}\times 2)+(\frac{1}{2}\times 2+\frac{1}{2}\times 2 + \frac{1}{2}\times 2)=8.5$.

\textbf{Downlink Communication:}
For the $z=2$ round, each relay needs to obtain $2\times |\mathcal{I}_2|=12$ segments, as each IV is divided into $2$ segments.
Note that the minimum number of IVs in $\mathcal{I}_2$ a relay has is $2$, i.e., $\min_{i\in[H]} |\mathcal{N}_i \cap \mathcal{I}_2|=2$. Hence the number of linear combinations needed to solve for all $12$ segments is $12- 2\times2=8$.
After receiving the uplink messages, the server generates $8$ random linear combinations of the receive messages $\{X_{i,j}^{ 2}: i\in[5], j\in[n_i^2]\}$, i.e.,
\begin{IEEEeqnarray}{rcl}
    X_{0,n}^{2}=C_{0,n}^{2} (X_{i,j}^{2}: i\in[5], j\in[n_i^2]),
\end{IEEEeqnarray}
for $n=1,\ldots,8$, where $\{C_{0,n}^{2}(\cdot)\}_{n=1}^8$ are random linear combining functions.

For the $z=3$ round, each relay needs to obtain $2\times |\mathcal{I}_3|=8$ segments, as each IV is actually divided into $2$ segments.
Note that the minimum number of IVs in $\mathcal{I}_3$ a relay has is $1$, i.e., $\min_{i\in[H]} |\mathcal{N}_i \cap \mathcal{I}_3|=1$. Hence the number of linear combinations needed to solve for all $8$ segments is $8- 2\times1=6$.
After receiving the uplink messages, the server generates $6$ random linear combinations of the receive messages $\{X_{i,j}^{3}: i\in\{2,4,5\}, j\in[n_i^3]\}$, i.e.,
\begin{IEEEeqnarray}{rcl}
    X_{0,n}^{3}=C_{0,n}^{3} (X_{i,j}^{3}: i\in\{2,4,5\}, j\in[n_i^3]),
\end{IEEEeqnarray}
for $n=1,\ldots,6$, where $\{C_{0,n}^{3}(\cdot)\}_{n=1}^6$ are random linear combining functions.

After finishing the $z\in\{2,3\}$ communication rounds, each relay can recover all the IVs, complete the global update, and transmit the latest model via the downlink to users.
Hence, using the coding methods, we have the downlink communication load
$L_{\textnormal{down}}^{S\rightarrow \mathcal{E}}=\frac{1}{2}\times 8 +\frac{1}{2}\times 6=7$. Therefore, the proposed scheme achieves the communication load pair $(8.5, 7)$, much smaller than the load pair $(K,K)=(10,10)$ achieved by the uncoded scheme.

\section{The Proposed Coded Scheme}\label{sec: scheme}
In this section, we present a novel coded scheme to reduce communication loads of the hierarchical MTL framework. Before the transmission, user $k\in [K]$ generates the IV $\mathbf{v}_k \in \mathbb{F}_{2^{V}}$ during the local update phase.

\textbf{Uplink Communication:}
User $k\in [K]$ first transmits the IV directly to the set of relays $\{i:k \in \mathcal{N}_i\}$ via the uplink, i.e., $X_k^{\mathcal{U}\rightarrow \mathcal{E}}=\mathbf{v}_k$. For the transmission between the relays and server, we  divide the transmission into $H-1$ rounds, and the $z$ communication round  aims to send IVs $\{\mathbf{v}_k: k\in \mathcal{I}_z\}$. 
We first consider the $z$ communication round  in the uplink, $z\in[H-1]$. Recall that $\mathcal{I}_z$ in \eqref{DefIz} denotes the indices of IVs which are available at $z$ relays and $\alpha^z_iV$  represents the number of bits of the single message sent by the relay $i\in[H]$ in the $z$ round. For each IV $\mathbf{v}_k$ with $k\in \mathcal{I}_z$, we divide it into $z$ disjoint segments, i.e.,
\begin{IEEEeqnarray}{rcl}
    \mathbf{v}_k=(\mathbf{v}_{k,i}: k\in \mathcal{N}_i \cap \mathcal{I}_z),
\end{IEEEeqnarray}
where the number of bits in $\mathbf{v}_{k,i}$ is $\alpha^z_i V$ for all $  k\in \mathcal{N}_i \cap \mathcal{I}_z$. We let the relay $i$ be responsible for sending all segments $\mathbf{v}_{k,i}$ with $k\in  \mathcal{N}_i \cap \mathcal{I}_z$ if $\alpha^z_i \neq 0$, and otherwise sending nothing.

Note that to ensure that each IV $\mathbf{v}_k$ with $k\in \mathcal{I}_z$ is transmitted successfully, the total number of  bits sent by the $z$ relays who can obtain $\mathbf{v}_k$ must be at least greater than $V$, i.e.,
\begin{IEEEeqnarray}{rcl}\label{equ: constraint1z}
    \sum_{i=1}^H \alpha^z_i \mathds{1}_{\mathcal{N}_i}(k) \ge 1, k\in \mathcal{I}_z,
\end{IEEEeqnarray}
where $\mathds{1}_{\mathcal{N}_i}$ is defined as an indicator function, i.e., $\mathds{1}_{\mathcal{N}_i}(k)=1$ if $k \in \mathcal{N}_i$ and $\mathds{1}_{\mathcal{N}_i}(k)=0$ if $k \notin \mathcal{N}_i$. 
For all $z\in[H-1]$ rounds of communication, from \eqref{equ: constraint1z}, we can obtain  the following   constraint
\begin{IEEEeqnarray}{rCl}
 \sum_{i=1}^H \alpha^z_i \mathds{1}_{\mathcal{N}_i}(k) \ge 1,  \forall k \in [K] \label{equ: constraint1}.
\end{IEEEeqnarray}
% in  (\ref{equ: constraint1}).   

In the $z$ round, we know that for each relay $i$,  the minimum number of common IVs  shared by  another relay   $j\in [H] \backslash \{i\}$ is $\min_{j \in [H]}|\mathcal{N}_i\cap \mathcal{N}_j \cap \mathcal{I}_z|$, i.e., there are at most $n_i^z=|\mathcal{N}_i \cap \mathcal{I}_z | - \min_{j \in [H]}|\mathcal{N}_i\cap \mathcal{N}_j \cap \mathcal{I}_z|$ IVs not known by the remaining relays $j\in [H] \backslash \{i\}$.
Thus $ n_i^z$ linearly independent combinations are able to help all other relays decode the unknown IVs. In the uplink communication of $z$ round, the relay $i \in [H]$ sends
\begin{IEEEeqnarray}{rcl} \label{equ: uplinkScheme}
    X_{i,j}^{z} = C_{i,j}^{ z}\left(\mathbf{v}_{k,i}: k\in \mathcal{I}_z \cap \mathcal{N}_i \right),
\end{IEEEeqnarray}
for $j=1,\ldots, n^z_i$, where $\{C_{i,j}^{z}(\cdot)\}_{i=1}^{n^z_i}$ are random linear combining functions. 
Hence the uplink communication load $L_{\textnormal{up},z}^{\mathcal{E}\rightarrow S}$ of the $z$ round  is
\begin{IEEEeqnarray}{rcl}
    L_{\textnormal{up},z}^{\mathcal{E}\rightarrow S}=\sum_{i=1}^H n_i^z \alpha^z_i.\label{equ: uplink}
\end{IEEEeqnarray}
Based on the closed-form of the communication load, we optimize the parameters $\{\alpha^z_i\}_{i\in [H]}$ related to the proposed transmission scheme, i.e, we formulate the optimization problem, 
\begin{align}
    \mathcal{P}(z): \label{equ: optimization_problem}\\
     \min_{\{\alpha^z_i\}_{i\in [H]}} &L_{\textnormal{up},z}^{\mathcal{E}\rightarrow S} \tag{\ref{equ: optimization_problem}{a}} \label{equ: objective}\\
    s.t.~&  \sum_{i=1}^H \alpha^z_i \mathds{1}_{\mathcal{N}_i}(k) \ge 1,  \forall k \in [K] \tag{\ref{equ: optimization_problem}{b}} \label{equ: constraint1}, \\ 
    &  0 \le \alpha^z_i\le 1, \forall i \in [H] \tag{\ref{equ: optimization_problem}{c}} \label{equ: constraint2}.
\end{align}

Note that the above optimization problem is a linear programming problem, and  can be solved efficiently (e.g., using interior point methods) \cite{boyd2004convex} with computational complexity $O(H^3)$ \cite{vaidya1987algorithm}. 

Based on the optimal $\{\alpha^z_i\}_{z\in[H], i\in [H]}$, we can derive the achievable uplink communication load,
\begin{IEEEeqnarray}{rcl}
    L_{\textnormal{up}}^{\mathcal{E}\rightarrow S}=
    \sum_{z=1}^{H-1} \sum_{i=1}^H  \alpha^z_i \left(
    \left| \mathcal{N}_i \cap \mathcal{I}_z  \right|  -\min_{j\in[H]} \left| \mathcal{N}_i \cap \mathcal{N}_j \cap \mathcal{I}_z  \right| \right).
\end{IEEEeqnarray}

\textbf{Downlink Communication:} Considering the $z$ communication round in the downlink, we first choose a parameter $0<\alpha^z \le 1$ such that $1/\alpha^z \in \mathbb{N}^+$ and ${\alpha^z_i}/{\alpha^z}\in \mathbb{N}$, $\forall i\in[H]$. Based on $\alpha^z$, the server divides the received uplink messages $\{X_{i,j}^{z}\}_{i\in[H], j\in [n^z_i]}$ into disjoint equal segments with bit number $\alpha^z V$. For each uplink message $X_{i,j}^{z}$, we have
\begin{IEEEeqnarray}{rcl}
    X_{i,j}^{z} =  \left(X_{i,j}^{z}[1], X_{i,j}^{z}[2],\ldots,X_{i,j}^{z}\left[\frac{\alpha^z_i}{\alpha^z}\right]  \right).
\end{IEEEeqnarray}
for $i=1,\ldots,H$ and $j=1,\ldots,n^z_i$.
The server generates $\frac{V}{\alpha^z}\left(|\mathcal{I}_z| -\min_{i\in[H]} |\mathcal{I}_z \cap  \mathcal{N}_i| \right)$ random linear combinations $X_{0,k}^{z}$ of the uplink message segments, i.e.,
\begin{IEEEeqnarray}{rcl}
    X_{0,n}^{z}= C_{0,n}^{z} \left(X_{i,j}^{z}[s]: i\in [H], j\in n^z_i, s\in \left[\frac{\alpha^z_i}{\alpha^z}\right]\right),
\end{IEEEeqnarray}
for $n=1,\ldots,\frac{V}{\alpha^z}\left(|\mathcal{I}_z| -\min_{i\in[H]} |\mathcal{I}_z\cap \mathcal{N}_i | \right)$, where $\{C_{0,n}^{z}(\cdot)\}_{n=1}^{\frac{V}{\alpha^z}|\mathcal{I}_z| -\min_{i\in[H]} |\mathcal{I}_z\cap \mathcal{N}_i |}$ are random linear combining functions. 

Note that each IV $\mathbf{v}_k$ with $k\in \mathcal{I}_z$ consists of $\frac{1}{\alpha^z}$ segments of bit number $\alpha^z V$, and hence every relay $i\in [H]$ needs  $\frac{1}{\alpha^z}|\mathcal{I}_z|$ segments and has already known $\frac{1}{\alpha^z}|\mathcal{I}_z\cap \mathcal{N}_i |$ segments. After receiving $\frac{1}{\alpha^z}\left(|\mathcal{I}_z| -\min_{i\in[H]} |\mathcal{I}_z\cap \mathcal{N}_i | \right)$ independent linear combinations $\{X_{0,n}^{z}\}$ sent by the server, based on the local IVs and the received linear combinations, the relay $i$ is able to decode the desired $\frac{1}{\alpha^z}\left(|\mathcal{I}_z|-|\mathcal{I}_z\cap \mathcal{N}_i |\right)$ segments because it has more independent linear combinations than its unknown segments, i.e.,
$$
\frac{1}{\alpha^z}\left(|\mathcal{I}_z|-|\mathcal{I}_z\cap \mathcal{N}_i |\right) \le \frac{1}{\alpha^z}\left( |\mathcal{I}_z| -\min_{i\in[H]} |\mathcal{I}_z\cap \mathcal{N}_i |\right).
$$

Finally, all relays obtain all IVs and apply the global update function in  (\ref{equ:global_update}), and then the relays send the updated model parameters $\mathbf{w}_k$ to user $k\in[K]$. Using the delivery strategies described above, the downlink communication load is
\begin{IEEEeqnarray}{rcl}
    L_{\textnormal{down}}^{S\rightarrow \mathcal{E}} &&= \sum_{z=1}^{H-1} \alpha^z V \cdot \frac{1}{\alpha^z}\left( |\mathcal{I}_z| -\min_{i\in[H]} |\mathcal{I}_z\cap \mathcal{N}_i |\right)\cdot \frac{1}{V} \nonumber \\
    &&= \sum_{z=1}^{H-1} \left(|\mathcal{I}_z| -\min_{i\in[H]} |\mathcal{I}_z\cap \mathcal{N}_i | \right) \nonumber \\
    &&= K- \sum_{z=1}^{H-1} \min_{i\in[H]} |\mathcal{I}_z\cap \mathcal{N}_i |.
\end{IEEEeqnarray}

Hence by the scheme described above, we obtain the following theorem.

\begin{theorem}\label{thm: upper} {For the hierarchical distributed MTL with the network connection matrix $G$, the corresponding network connection indices $\{\mathcal{N}_i\}_{i\in[H]}$ and  $z$-connected users $\{I_z\}_{z\in[H]}$, the communication loads $(L_\textnormal{up}^{\mathcal{E}\rightarrow S}, L_\textnormal{down}^{S\rightarrow \mathcal{E}})$ are achievable,} where
\begin{subequations}\label{fixeload}
\begin{IEEEeqnarray}{rcl}
    &&L_{\textnormal{up}}^{\mathcal{E}\rightarrow S}= 
    \sum_{z=1}^{H-1} \sum_{i=1}^H  \alpha^z_i \left(
    \left| \mathcal{N}_i \cap \mathcal{I}_z  \right|  -\min_{j\in[H]} \left| \mathcal{N}_i \cap \mathcal{N}_j \cap \mathcal{I}_z  \right| \right),\label{equ: up}\\ 
    &&L_{\textnormal{down}}^{S\rightarrow \mathcal{E}}=
    K-\sum_{z=1}^{H-1} \min_{i\in [H]}\left|\mathcal{N}_i \cap I_z\right|,\label{equ: down}
\end{IEEEeqnarray}
\end{subequations}
where the choice of $\mathcal{\alpha}_z^i$ is determined by solving the optimization problem in (\ref{equ: optimization_problem}).
\end{theorem}

\begin{figure}[htbp]
	\centering
        \subfigure[Uplink Communication Loads]{
            \centering
		\includegraphics[scale=0.5]{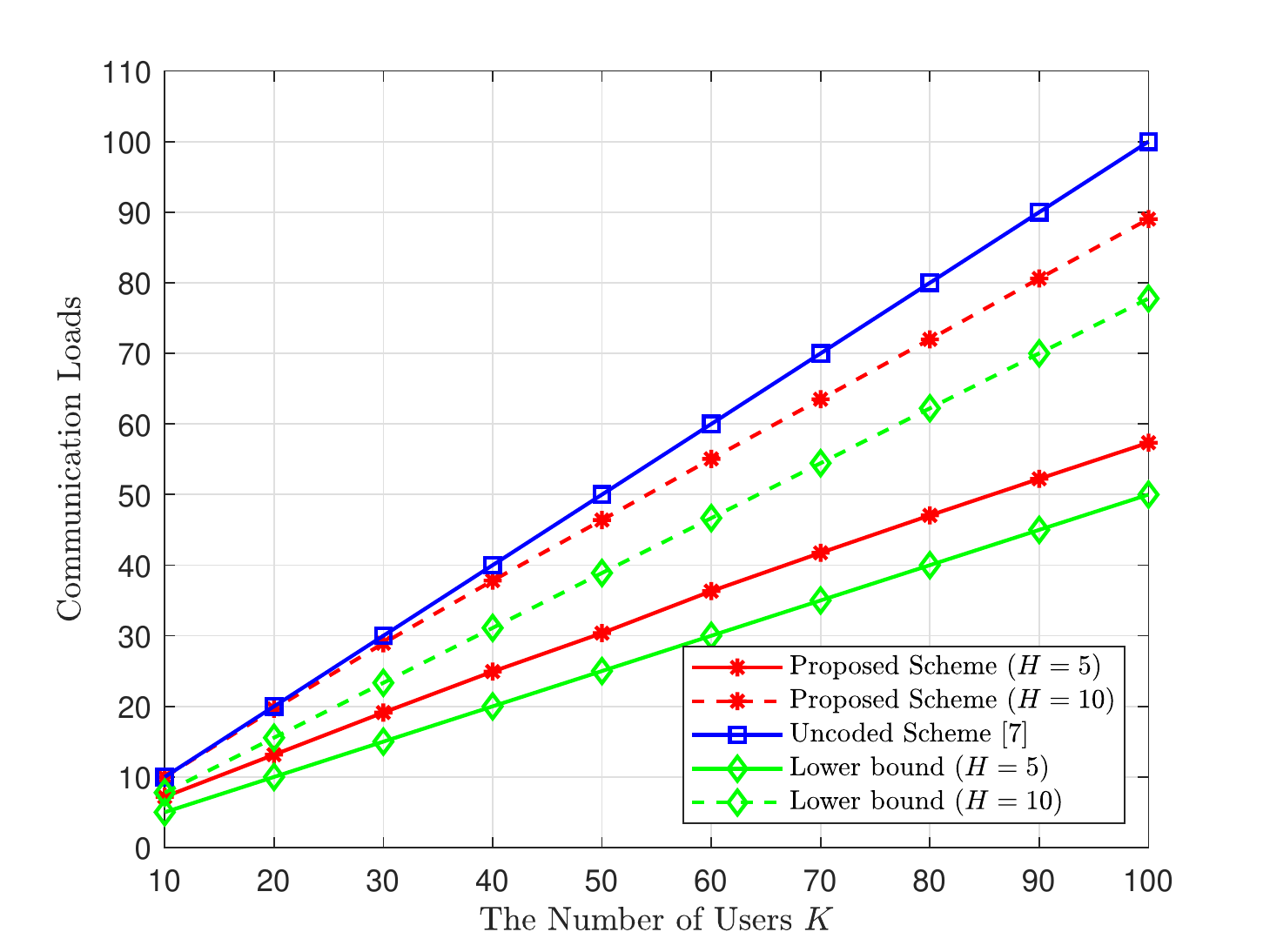}
        }
        \subfigure[Downlink Communication Loads]{
            \centering
		\includegraphics[scale=0.5]{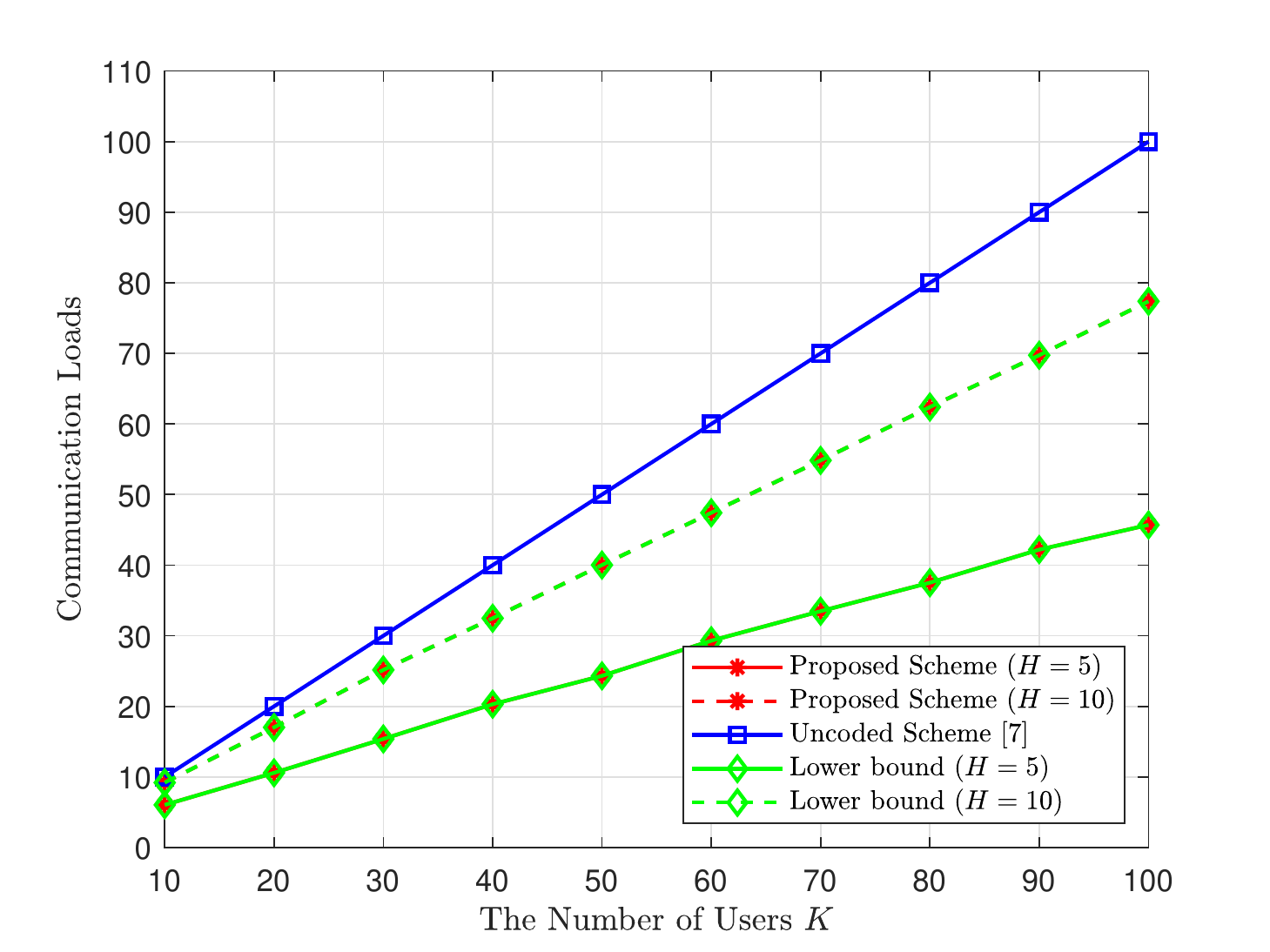}
        }
	\centering
	\caption{Comparison of the communication loads with different $K$, $H\in \{5, 10\}$ relays, and each user connects to any three relays: (a) Uplink communication loads; (b) Downlink communication loads.}
	\label{fig: loads_different_K}
\end{figure}

\begin{Remark} \label{remark: flex}
    The constraints of the optimization problem (\ref{equ: optimization_problem}) can be satisfied by using an equal partitioning approach, i.e., each relay $i$ that can obtain IVs $\mathbf{v}_k$ sends $\frac{V}{z}$  bits of $\mathbf{v}_k$, $k\in \mathcal{N}_i$. Using this method, we have $\alpha^z_i =\frac{1}{z}$ if $|\mathcal{N}_i \cap \mathcal{I}_z| \neq 0$ and $\alpha^z_i = 0$ if $|\mathcal{N}_i \cap \mathcal{I}_z| = 0$, and the following upper bound of $L_\textnormal{up}$ can be derived: 
%    \begin{align}
        $L_{\textnormal{up}}^{\mathcal{E}\rightarrow S}\le K- \sum_{z=1}^{H-1} \sum_{i=1}^H  \frac{1}{z} \min_{j\in[H]} \left| \mathcal{N}_i \cap \mathcal{N}_j \cap \mathcal{I}_z  \right|.$
   % \end{align}
\end{Remark}

\begin{Remark}
    From Theorem \ref{thm: upper} and Remark \ref{remark: flex},  the communication load reduction between the proposed scheme and the uncoded scheme is as follows,
    \begin{subequations}\label{equ: reduction}
    \begin{IEEEeqnarray}{rcl}
        L_{\textnormal{up,uncoded}}^{\mathcal{E}\rightarrow S}-L_{\textnormal{up}}^{\mathcal{E}\rightarrow S} &&\ge \sum_{z=1}^{H-1} \sum_{i=1}^H  \frac{1}{z} \min_{j\in[H]} \left| \mathcal{N}_i \cap \mathcal{N}_j \cap \mathcal{I}_z  \right|,\\ 
        L_{\textnormal{down,uncoded}}^{S\rightarrow \mathcal{E}}-L_{\textnormal{down}}^{S\rightarrow \mathcal{E}} &&=
    \sum_{z=1}^{H-1} \min_{i\in [H]}\left|\mathcal{N}_i \cap I_z\right|.
    \end{IEEEeqnarray}
\end{subequations}
\end{Remark}
Compared with the communication load pair of the uncoded scheme, the proposed scheme can significantly reduce the communication loads both in the uplink and downlink.
The lower bounds of communication loads for the hierarchical distributed MTL with fixed network connection are given in the following theorem.

\begin{Remark}
    Note that our scheme is highly adaptable and can fit into any network connection between users and relays.
    This is quite different from previous coded caching problems, which mostly allowed flexible data placement at relays or users and considered special network typologies such as combinatorial networks \cite{ji2015fundamental,tang2016coded,zewail2017coded,zewail2019cache,zewail2018combination}.
\end{Remark}

\begin{figure}[htbp]
	\centering
        \subfigure[Uplink Communication Loads]{
            \centering
		\includegraphics[scale=0.5]{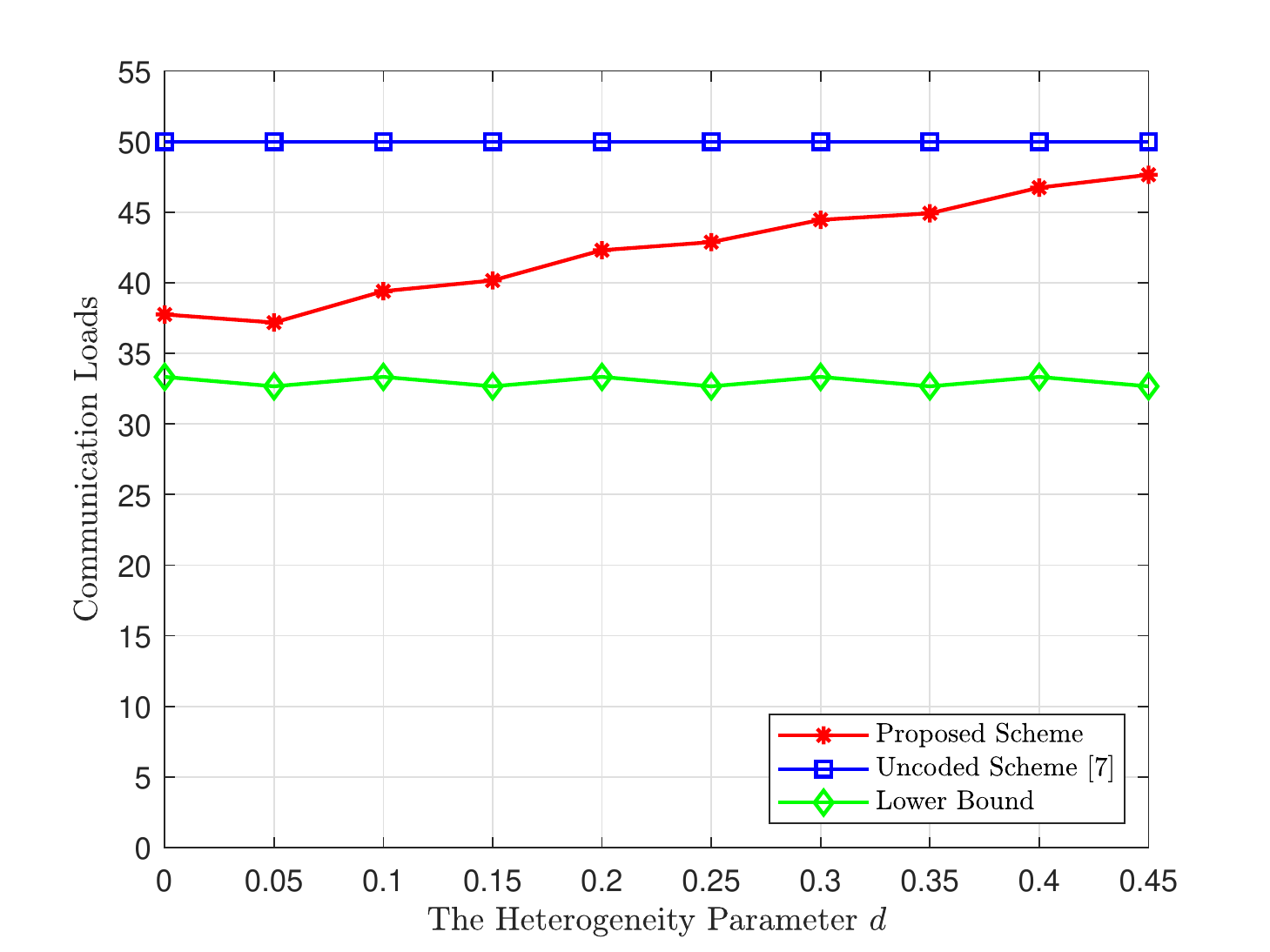}
        }
        \subfigure[Downlink Communication Loads]{
            \centering
		\includegraphics[scale=0.5]{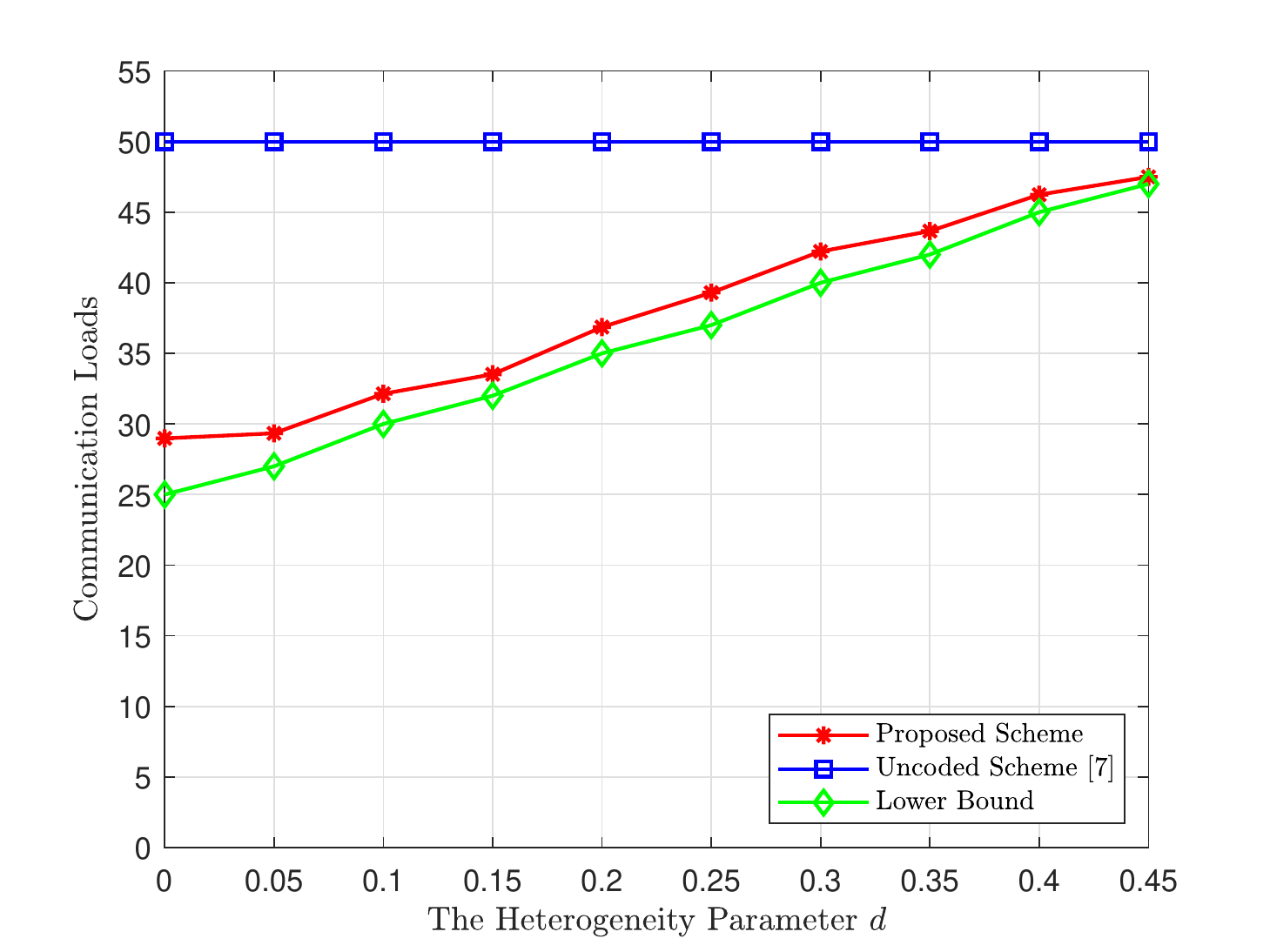}
        }
	\centering
	\caption{Comparison of the  communication loads with $K=50$ users, $H=4$ relays, and different heterogeneity parameters $d$: (a) Uplink communication loads; (b) Downlink communication loads.}
        \label{fig: loads_different_d}
\end{figure}

\begin{theorem}\label{thm: lower} For the hierarchical distributed MTL with the network connection matrix $G$, the corresponding network connection indices $\{\mathcal{N}_i\}_{i\in[H]}$ and  $z$-connected users $\{I_z\}_{z\in[H]}$, the optimal communication loads $(L_\textnormal{up}^{\mathcal{E}\rightarrow S*}, L_\textnormal{down}^{S\rightarrow \mathcal{E}*})$ satisfy
\begin{subequations}
\begin{IEEEeqnarray}{rCl} 
&&L_\textnormal{up}^{\mathcal{E}\rightarrow S*}\geq \frac{HK}{H-1}-\frac{\sum_{z=1}^{H} z|\mathcal{I}_z|}{H-1}, \label{equ: upLower}\\ 
&&L_\textnormal{down}^{S\rightarrow \mathcal{E}*}\geq {K}-\min_{i\in[H]}|\mathcal{N}_i| \label{equ: downLower}.
\end{IEEEeqnarray}
\end{subequations}
\end{theorem}

\begin{IEEEproof}
See Appendix \ref{sec: Opt}.
\end{IEEEproof}

\begin{subequations} 
\begin{Corollary}\label{corollary: gap} For the hierarchical distributed MTL with the network connection matrix $G$, the corresponding network connection indices $\{\mathcal{N}_i\}_{i\in[H]}$ and  $z$-connected users $\{I_z\}_{z\in[H]}$, we have 
\begin{IEEEeqnarray}{rcl}
L_{\textnormal{up}}^{\mathcal{E}\rightarrow S}-L_{\textnormal{up}}^{\mathcal{E}\rightarrow S*} &&\le \min_{i\in[H]}|\mathcal{N}_i| - \sum_{z=1}^{H-1} \sum_{i=1}^H  \frac{1}{z} \min_{j\in[H]} \left| \mathcal{N}_i \cap \mathcal{N}_j \cap \mathcal{I}_z \right| \nonumber \\
&& \le \min_{i\in[H]} |\mathcal{N}_i|, \\
L_{\textnormal{down}}^{S\rightarrow \mathcal{E}}-L_{\textnormal{down}}^{S\rightarrow \mathcal{E}*} &&\le \min_{i\in[H]} |\mathcal{N}_i| - \sum_{z=1}^{H-1} \min_{i\in [H]}\left|\mathcal{N}_i \cap I_z\right| \nonumber \\
&& \le \min_{i\in[H]} |\mathcal{N}_i|. 
\end{IEEEeqnarray}

\end{Corollary}
\end{subequations}

\begin{IEEEproof}
    See Appendix \ref{sec: gap}.
\end{IEEEproof}

\begin{Remark}
    Note that the gaps between achieved upper bounds and lower bounds are within  the minimum number of connected users among all relays both in the uplink and downlink.
    This demonstrates the scalability of our schemes, which means that they can be used with a large number of users.
\end{Remark}

\begin{theorem}\label{corallary: flexible}
Given average user connectivity $r$, we assume that the network connection matrix can be delicately designed. Under this setting, using a symmetric design (similar to the network topology of a combination network\cite{ngai2004network,zewail2018combination}) and using the above method, the optimal communication load can be achieved, \begin{subequations}\begin{IEEEeqnarray}{rcl}
L_{\textnormal{up}}^{\mathcal{E}\rightarrow S}&&= L_{\textnormal{up}}^{\mathcal{E}\rightarrow S*}= \frac{K(H-r)}{H-1}, \\
L_{\textnormal{down}}^{S\rightarrow \mathcal{E}}&&= L_{\textnormal{down}}^{S\rightarrow \mathcal{E}*}= K-\frac{rK}{H}, 
\end{IEEEeqnarray}
and the achievable load pair is optimal, i.e., our scheme achieves the minimum load pair both in the downlink and uplink communications. %  the optimum.
\end{subequations}
\end{theorem}

\begin{IEEEproof}
    See Appendix \ref{sec: flex}.
\end{IEEEproof}

From Theorem \ref{corallary: flexible}, we learn that our proposed scheme can be directly applied to scenarios where the network topology can be designed flexibly, and the communication loads achieve optimal, which illustrates the superiority of our scheme.

The numerical result of the achievable communication load pair $(L_\textnormal{up}^{\mathcal{E}\rightarrow S}, L_\textnormal{down}^{S\rightarrow \mathcal{E}})$ is presented in Fig. \ref{fig: loads_different_K} and Fig. \ref{fig: loads_different_d}. 
We first consider the setting with different $K$ users  and $H\in \{5,10\}$ relays, and each user connects to any three relays.
As is shown in Fig. \ref{fig: loads_different_K}, both the uplink and downlink communication load  of our scheme (the red star line) are much smaller than those of the uncoded scheme  (the blue square line). 
Besides, the higher the number of users, the more effective our scheme is in reducing communication loads.
This can be explained by the fact that more users bring more IVs overlap on relays, i.e. more side information, leading to a greater reduction in communication loads. The gap between the achievable scheme and the lower bound (the green rhombus line) is approximately the same for different $K$, further illustrating the scalability of our scheme under large-scale networks.
In addition, the communication loads with $H=5$ (the red star solid line) is smaller than those with $H=10$ (the red star dotted line).
This is due to the fact that for a given average user connectivity $r$, the higher the number of relays $H$, the lower the number of IVs obtained per relay and the lower the coding opportunities.
Moreover, our scheme is more effective in reducing the downlink communication load than the uplink, where the coded scheme is almost information-theoretic optimal for the downlink in the simulation.

Next, we consider a setting with $K=50$ users, $H=4$ relays, and the total number of links between users and relays is $2K=100$. We assume that the number of users connected by  each relay is $|\mathcal{N}_1|=|\mathcal{N}_2|= (\frac{1}{2}+d) \cdot K$ and $|\mathcal{N}_3|=|\mathcal{N}_4|= (\frac{1}{2}-d) \cdot K$ with the heterogeneity parameter $d$, and relay $i$ connects any $|\mathcal{N}_i|$ users. The parameter $d$ represents the degree of heterogeneity of the distributed system, as when $d$ grows larger, the difference in the number of relay connections becomes larger, i.e. the system is more heterogeneous.
As is shown in Fig. \ref{fig: loads_different_d}, our proposed scheme  (the red star line) outperforms the uncoded scheme (the blue square line) with all $d$.
Especially, when $d$ is smaller, the superiority of the  proposed scheme is more obvious.
The achieved upper bound converges to the uncoded scheme when $d$ grows, as there are few coding opportunities to reduce communication loads under highly heterogeneous scenarios. Similarly, our proposed scheme reduces the downlink communication load more significantly than the uplink.

\section{Experiments}\label{sec: experiement}
In this section, we apply our proposed coded scheme in Section \ref{sec: scheme} to the MOCHA algorithm \cite{smith2017federated}, and demonstrate our superiority in comparison with the uncoded scheme.
Note that our scheme allows for lossless data transmission, thus the scheme can be combined with other data compression schemes, i.e, sparsification \cite{aji2017sparse,stich2018sparsified,wangni2017gradient}, quantization \cite{alistarh2017qsgd,bernstein2018signsgd,liang2021improved}.
Hence we do not consider the comparison with other lossy compression schemes in our experiment. 
In addition, since our scheme has the same training performance and convergence rate as the original MOCHA scheme, we only consider the total training time as the experiment metric, which is calculated according to (\ref{equ: total}) and (\ref{equ: comm}).  

The MOCHA algorithm is a prevalent optimization algorithm for solving the MTL problem in (\ref{eq1}), which uses a primal-dual formulation to optimize the learned models. At each iteration, the distributed users perform the local update on data-local sub-problems and send generated IVs to the central server. After receiving IVs from all the users, the server executes the global update and sends different updated parameters to each user. Note MOCHA algorithm is consistent with the system model introduced in Section \ref{sec: system_model}, hence we can apply the proposed coded scheme with the MOCHA learning algorithm. 
We extend MOCHA to a hierarchical network, where users train based on local datasets to obtain IVs and send them in the uplink, with the final goal of obtaining a globally updated model from relays.
We use the uncoded scheme in Example \ref{example: uncoded} as the comparison scheme, and the communication load pair $(L_\textnormal{up,uncoded}^{\mathcal{U}\rightarrow \mathcal{E}}, L_\textnormal{up,uncoded}^{\mathcal{E}\rightarrow S}, L_\textnormal{down,uncoded}^{S\rightarrow \mathcal{E}}, L_\textnormal{down,uncoded}^{\mathcal{E}\rightarrow \mathcal{U}})=(K,K,K,K)$.

We choose an experimental setup similar to that in \cite{smith2017federated}, with the specific experimental details described below.  

\subsection{Experiment Setting and Datasets}
In our experiments, we select the hinge loss function as the loss function, and the best regularization parameter is selected  from $\{$1e-5, 1e-4, 1e-3, 1e-2, 0.1, 1, 10$\}$, for each model using 5-fold cross-validation. 
In the local update phase, we set each user to train 150 rounds, and at each iteration, we select 50\% local data for training. We perform 64-bit quantization in every communication phase and set the communication bandwidth $W_1=W_2=100$ Mbps. Each user is equipped with an Intel Core i7-9750H CPU with 16G RAM, and a working frequency 2.60GHz. For a fair comparison, we apply the same local update rule and global update rules as \cite{smith2017federated}, where users use an SVM to train local models based on local data and perform classification tasks.
To reduce the randomness of the experiment, for each experimental setup, we generate 50 random data placements and average their results as experimental results.
The optimization problem in (\ref{equ: optimization_problem}) is solved by using the optimization solver CVX \cite{grant2014cvx}.
The datasets in our experiments are as follows.
\begin{itemize}
    \item \textbf{MNIST dataset:} 
    The MNIST dataset is a hand-written digit dataset, and the dimension of each data instance in MNIST $l=784$. We divide the dataset into several sets, and  each set contains 500 data instances. 
    The $k$-th set contains 500 data instances, 250 of them labeled with digit Mod$(k,10)-1$, and 250 with  random digits. We randomly split the data into 75\% training and 25\% testing.
    Each user $k\in [K]$ aims to classify digit Mod$(k,10)-1$ with other digits based on the $k$-th set. 
    \item \textbf{Human Activity Recognition dataset:} The Human Activity Recognition dataset is the 3-axial linear acceleration and 3-axial angular velocity dataset collected from 30 individuals when they perform one of six activities: {walking, walking-upstairs, walking-downstairs, sitting, standing, and lying-down}. 
    The dimension of each data instance in the dataset $l=561$.
    We divide the dataset into several sets, and each set contains 500 data instances. We randomly split the data into 75\% training and 25\% testing.
    Each user $k\in [K]$ aims to classify sitting with the other activities based on the $k$-th set. 
\end{itemize}

\begin{figure}[htbp]
	\centering
        \subfigure[]{
            \centering
		\includegraphics[scale=0.5]{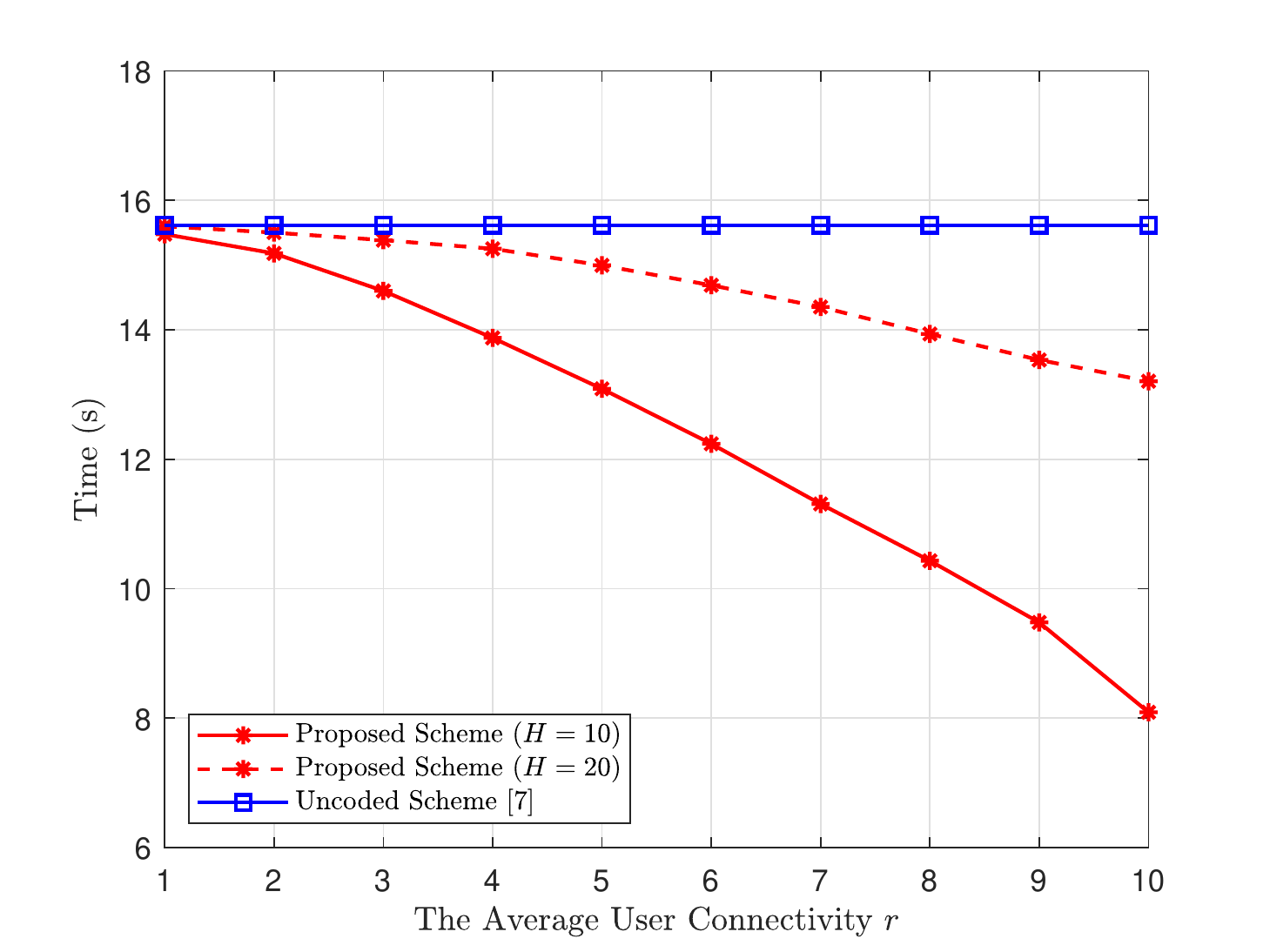}
        }
        \subfigure[]{
            \centering
		\includegraphics[scale=0.5]{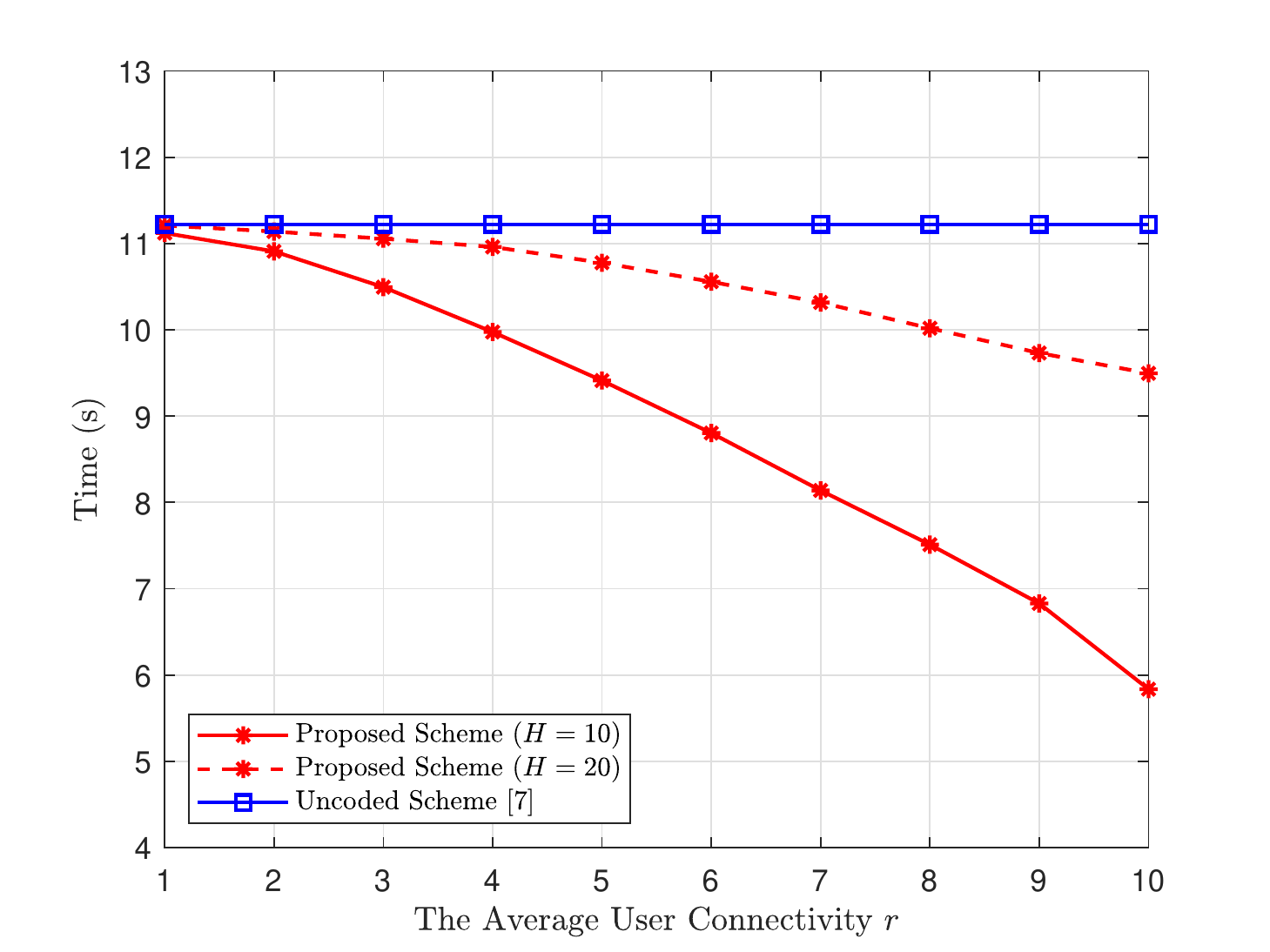}
        }
	\centering
	\caption{Comparison of the overall time with $K=20$ users, $H\in\{10, 20\}$ relays, and different average user connectivity $r$ using  (a) the MNIST dataset and (b) the Human Activity Recognition dataset.}
	\label{fig: different_r}
\end{figure}

\subsection{Experiment Results}
In the experiments, we compare the total execution time of our proposed scheme with the uncoded scheme. 
We first assume that each user has the same number of relay connections, i.e., $r_1=r_2\cdots=r_K=r$. We set the number of users $K=20$, and the number of relays $H\in\{10, 20\}$ while changing the average user connectivity $r\in[10]$, and the result is shown in Fig. \ref{fig: different_r}. 
Note that the total time of the uncoded scheme remains unchanged when the average user connectivity $r$ changes as we assume that the uncoded scheme
adopts the same delivery strategy in Section \ref{sec: system_model} for all $r$. 
As is shown in Fig. \ref{fig: different_r}, it is obvious that the total time of our scheme (the red star lines) is much smaller than that of the uncoded scheme (the blue square line). 
As $r$ increases, the total time proposed scheme decreases, which coincides with our analysis that more available IVs on relays can lead to more coding and multicast opportunities.
In the actual distributed system, we can make the user connect as many relays as possible to improve the system performance.
For different numbers of relays $H$, we note that the total time spent with $H=10$ (the red star solid line) is less than the total time spent with $H=20$ (the red star dotted line).
This is due to the fact that for a given average user connectivity $r$, the more the number of relays, the lower the number of IVs obtained per relay, and the coding opportunities decrease.
The observation guides us that the number of relays may not be as large as it could be when designing a hierarchical system.
We obtain the same trend of the curve both for the  MNIST dataset and the Human Activity Recognition dataset.

\begin{figure}[htbp]
	\centering
        \subfigure[]{
            \centering
		\includegraphics[scale=0.5]{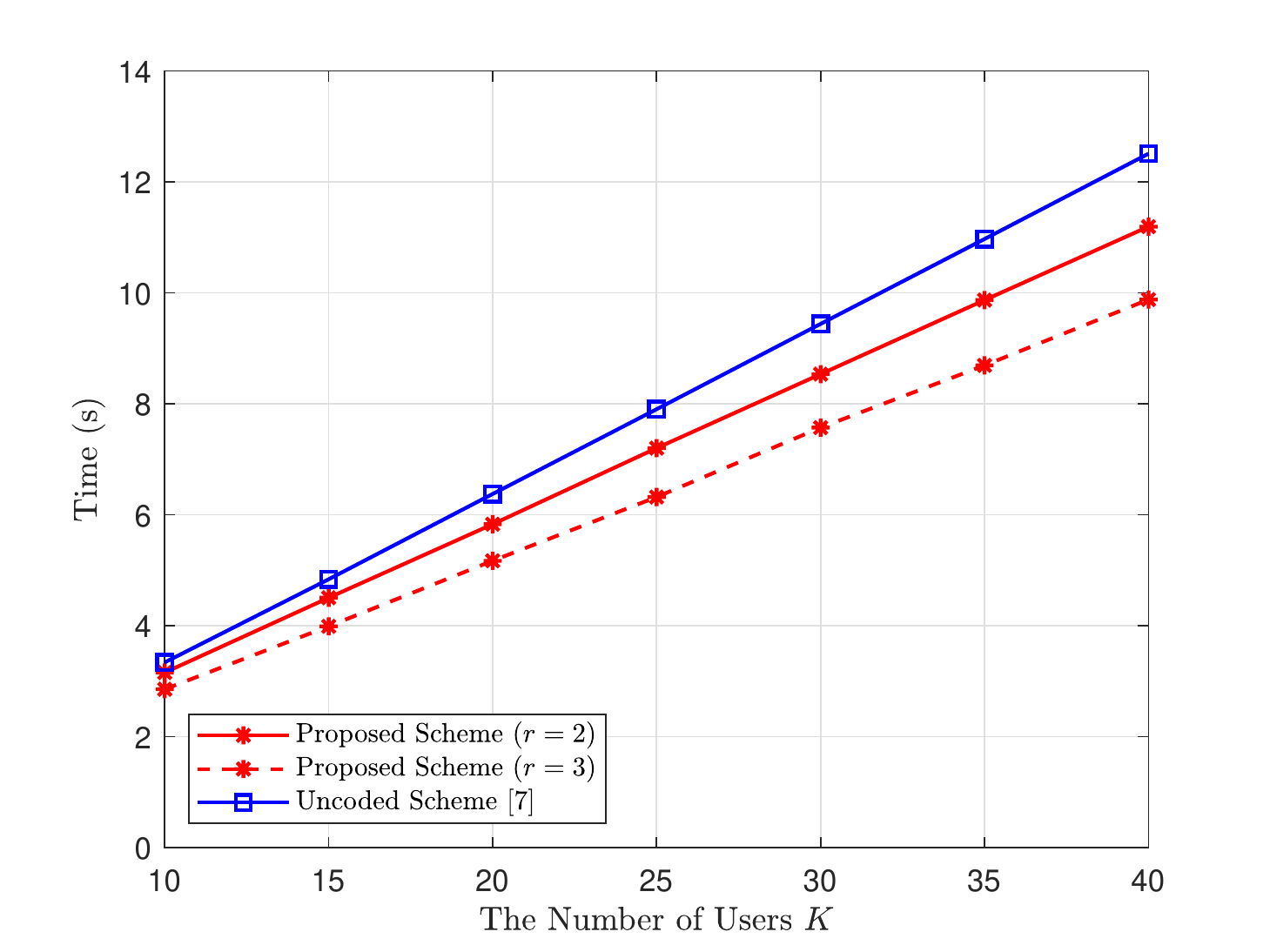}
        }
        \subfigure[]{
            \centering
		\includegraphics[scale=0.5]{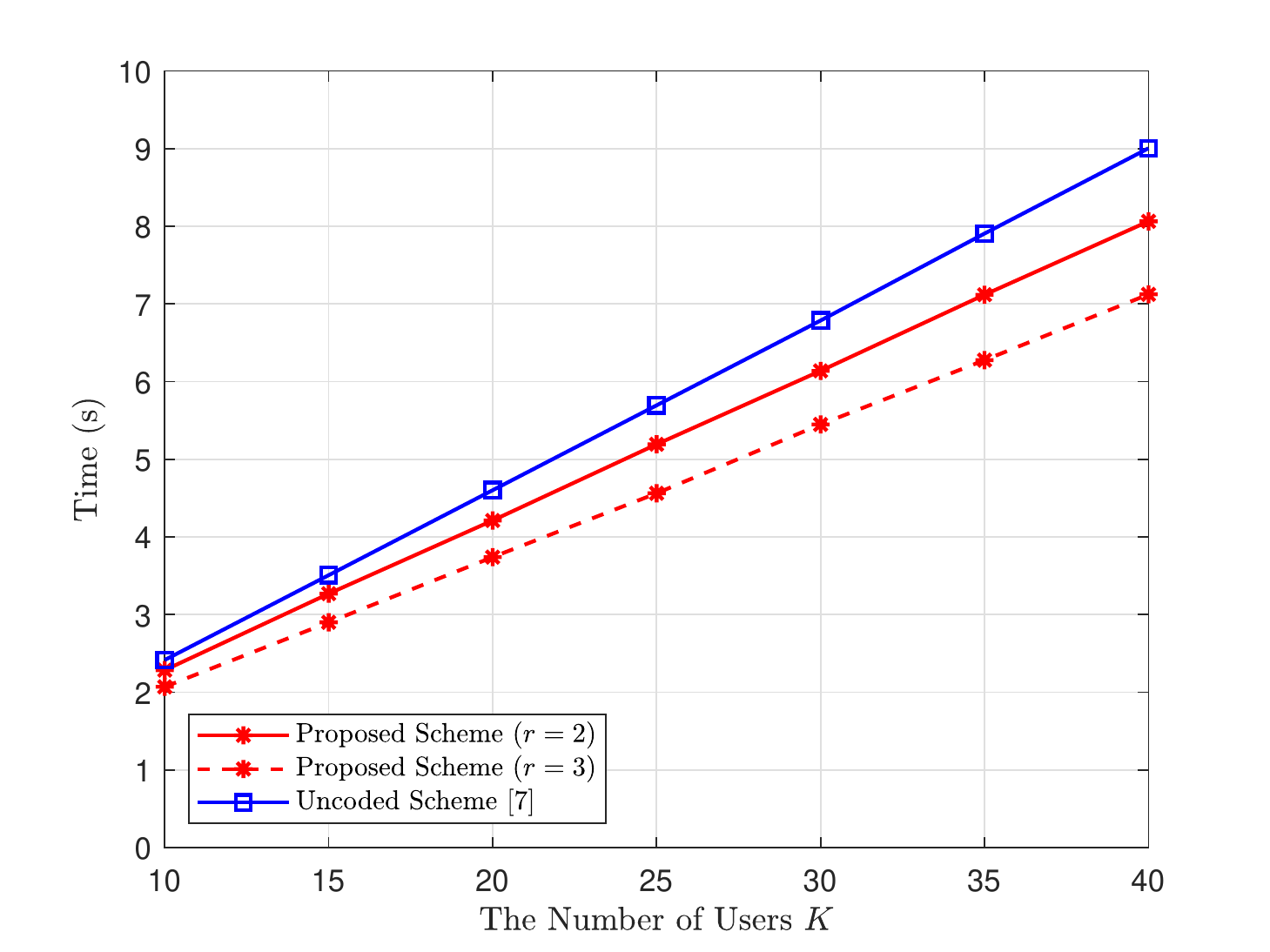}
        }
	\centering
	\caption{Comparison of total time with  $H=5$ relays, average user connectivity $r\in\{2, 3\}$, and  different numbers of users $K$ using (a)  the MNIST dataset and (b) the Human Activity Recognition dataset.}
	\label{fig: different_K}
\end{figure}

We then consider the experimental setting with different numbers of users $K$ and fixed average user connectivity $r$. We consider the setting that there are $H=5$ relays given average user connectivity $r\in\{2, 3\}$, and the number of users $K\in\{10,15,20,25,30,35,40\}$. 
We let each user connect to any $r$ relays, i.e., $r_1=r_2\cdots=r_K=r$.
As is shown in Fig. \ref{fig: different_K}, the total time of our scheme (the red star lines) is much smaller than that of the uncoded scheme (the blue square line). 
We note that the higher the number of users, the more effective our scheme is in reducing communication time.
In particular, for the MNIST dataset with $r=3$, the total time is reduced by about 17\% with $K=10$, while the total time is reduced by about 26\% with $K=40$. 
Moreover, the total time at $r=3$ (the red star dotted line) is lower than the total time at $r=2$ (the red star solid line), due to more connections creating more side information across the relays.
The trends of the curve for the  MNIST dataset and the Human Activity Recognition dataset are similar in our experiments.

\begin{figure}[htbp]
	\centering
        \subfigure[]{
            \centering
		\includegraphics[scale=0.5]{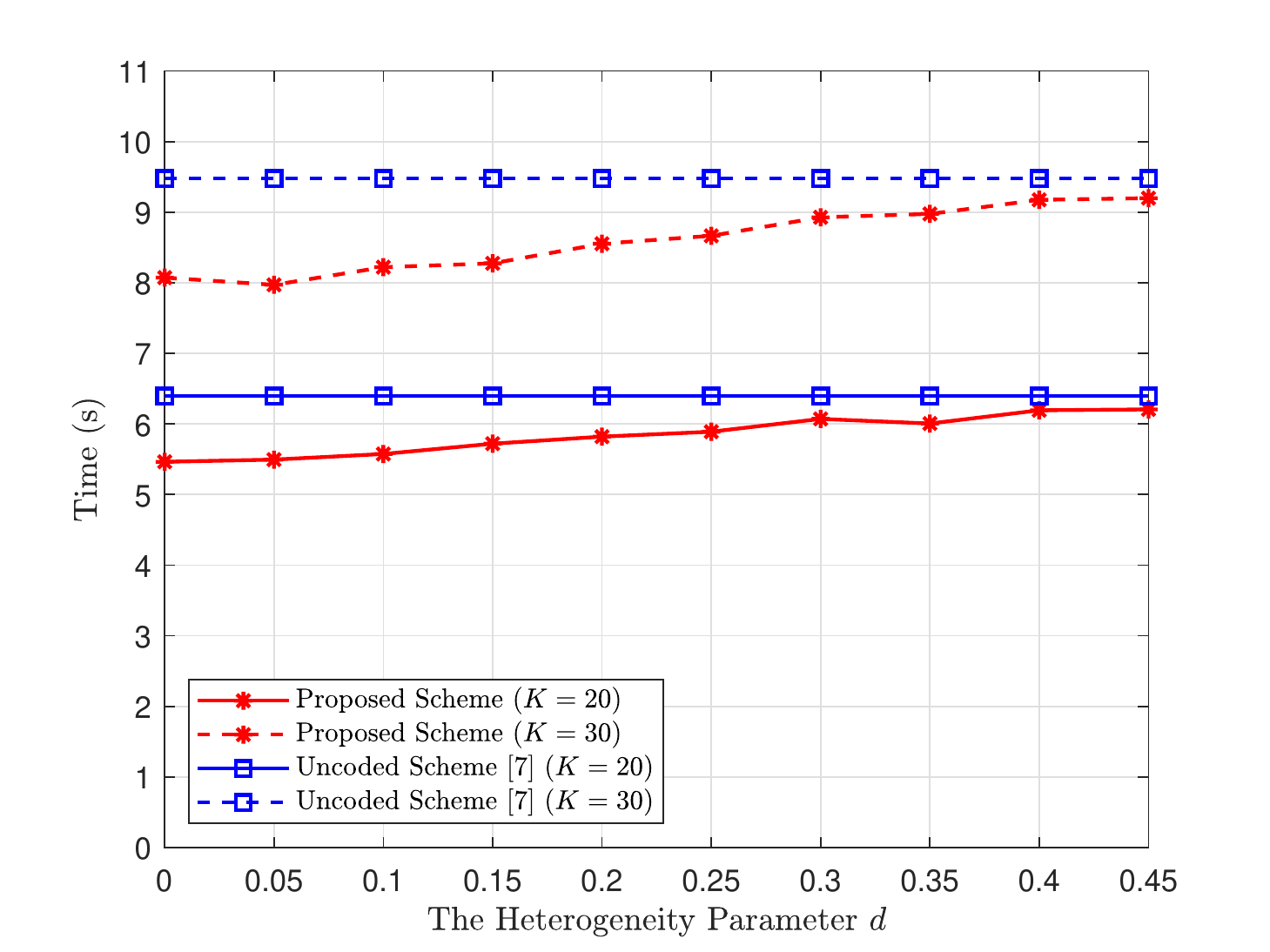}
        }
        \subfigure[]{
            \centering
		\includegraphics[scale=0.5]{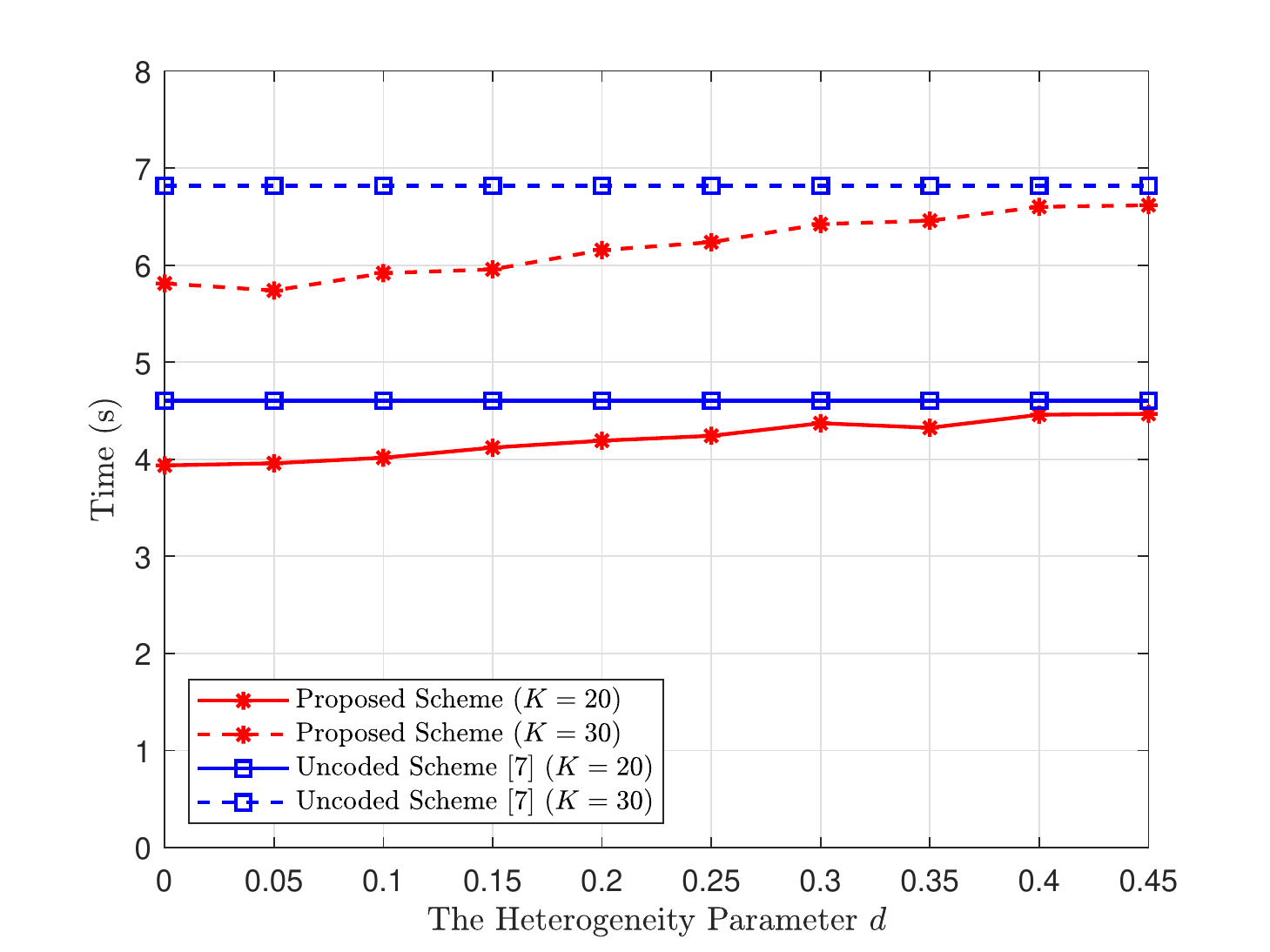}
        }
	\centering
	\caption{Comparison of total time with $K\in \{20, 30\}$ users, $H=4$ relays, average user connectivity $r=2$ and different heterogeneity parameters $d$ using (a) the MNIST dataset and (b) the Human Activity Recognition dataset.}
	\label{fig: different_d}
\end{figure}

Next, we consider a hierarchical distributed computing system with $K\in \{20, 30\}$ users, $H=4$ relays, and given average user connectivity $r=2$. Hence the total number of links between users and relays is $rK$, and we assume that the number of users connected by  each relay is $|\mathcal{N}_1|=|\mathcal{N}_2|= (\frac{r}{4}+d) \cdot K$ and $|\mathcal{N}_3|=|\mathcal{N}_4|= (\frac{r}{4}-d) \cdot K$, with heterogeneity parameter $d\in \{0,\frac{1}{20}, \frac{1}{10}, \ldots,\frac{9}{20} \}$. 
The experimental result is shown in Fig. \ref{fig: different_d}, and our proposed scheme (the red star line) achieves less time than the uncoded scheme (the blue square line) for all choices of $d$. The smaller the $d$, the less time our scheme achieves, which indicates that our solution is more suitable for symmetrical scenarios. 
In addition, the increase in the number of users can better reduce the total time, coinciding with our previous analysis.
For instance, for the  MNIST dataset with $K=30$, the total time is reduced by 17\% with $d=0$ while 3\% when $d=0.45$.
We obtain a similar trend of the curve with both experiment datasets.

\section{Conclusion}\label{sec: conclusion}
In this paper, we investigated the communication bottleneck of  multi-task learning problems under a hierarchical distributed computing system. We proposed a coded scheme to reduce the communication loads both in the uplink and downlink using the side information introduced on the relays.  We derived information-theoretic lower bounds of the communication loads for the hierarchical settings, and showed the gaps between our achievable communication loads and the optimum are within the minimum number of available IVs among all relays. Experiments on real-world datasets showed that the proposed scheme can greatly reduce the communication loads compared to the state-of-art approaches.
In future work, we would consider the wireless setting and search protocols that meet stronger privacy restrictions.

\appendices
\section{Proofs of Theorem \ref{thm: lower} and Corollary \ref{corollary: gap}}
\subsection{Proof of Theorem \ref{thm: lower}} \label{sec: Opt}
To prove the lower bound on $L_\textnormal{up}^{\mathcal{E}\rightarrow S*}$ and $L_\textnormal{down}^{S\rightarrow \mathcal{E}*}$ in Theorem \ref{thm: lower}, we first present a lemma proved in \cite{yu2017optimally}.

Define $a^\mathcal{Q}_\mathcal{P}$ as the number of IVs which are available at nodes in $\mathcal{Q}$ and required by (but not available at) nodes in $\mathcal{P}$, where $\mathcal{Q}\subseteq[K]$, $\mathcal{P}\subseteq [K]$. 

\begin{lemma}\label{Lemma1}
Consider a MapReduce-type task and a given Map and Reduce design that runs in a distributed computing system consisting of $K$ computing nodes. For any integers $q$, $p$, let $a_{q,p}$ denote the number of IVs that are available at $q$ nodes, and required by (but not available at) $p$ nodes. The following lower bound on the communication load holds, \begin{IEEEeqnarray}{rcl}  
    L\geq {\sum_{q=1}^{K} \sum_{p=1}^{K-q} a_{q,p} \frac{p}{q+p-1}}.
\end{IEEEeqnarray}
\end{lemma}

Under fixed network  connection, we have 
\begin{subequations}
\begin{IEEEeqnarray}{rcl} 
    &&\sum_{q=1}^{H}\sum_{\substack{\mathcal{Q}\subseteq[H],\\|\mathcal{Q}|=q}}a^\mathcal{Q}_{[K]\backslash \mathcal{Q}}=K, \label{constraint5} \\
    &&\sum_{q=1}^{H}\sum_{\substack{\mathcal{Q}\subseteq[H],\\|\mathcal{Q}|=q}}q a^\mathcal{Q}_{[H]\backslash \mathcal{Q}}=\sum_{q=1}^{H} q|\mathcal{I}_q|. \label{constraint6}
\end{IEEEeqnarray}
\end{subequations}

According to Lemma \ref{Lemma1}, for the uplink communication, we have  
\begin{IEEEeqnarray}{rcl}\label{formula_fix_up}
 L_\textnormal{up}^{\mathcal{E}\rightarrow S*}
 &&\ge \sum_{q=1}^{H} \sum_{p=1}^{H-q} a_{q,p} \frac{p}{q+p-1}\nonumber\\
 &&=\sum_{q=1}^{H} \sum_{p=1}^{H-q} \sum_{\substack{\mathcal{Q}\subseteq[H],\\|\mathcal{Q}|=q}} \sum_{\substack{\mathcal{P}\subseteq[H]\backslash\mathcal{Q},\\|\mathcal{P}|=p}} a^\mathcal{Q}_\mathcal{P} \frac{|\mathcal{P}|}{|\mathcal{P}|+|\mathcal{Q}|-1}\nonumber\\
 &&\overset{(a)}{=} \sum_{q=1}^{H} \sum_{\substack{\mathcal{Q}\subseteq[H],\\|\mathcal{Q}|=q}} a^\mathcal{Q}_{[H]\backslash \mathcal{Q}} \frac{H-q}{H-1}\nonumber\\
 &&=\sum_{q=1}^{H}\sum_{\substack{\mathcal{Q}\subseteq[H],\\|\mathcal{Q}|=q}} \frac{a^\mathcal{Q}_{[H]\backslash \mathcal{Q}}H}{H-1}-\sum_{q=1}^{H}\sum_{\substack{\mathcal{Q}\subseteq[H],\\|\mathcal{Q}|=q}} \frac{q a^\mathcal{Q}_{[H]\backslash \mathcal{Q}}}{H-1}\nonumber\\
 &&\overset{(b)}{=}\frac{HK}{H-1}-\frac{\sum_{q=1}^{H} q|\mathcal{I}_q|}{H-1}, 
\end{IEEEeqnarray}
where $(a)$ is due to the fact that each IV available at nodes in $\mathcal{Q}$ will be required by all the other nodes, i.e., $\mathcal{P}=[H]\backslash\mathcal{Q}$, and $(b)$ is due to the constraint of $\{a^\mathcal{Q}_{[H]\backslash \mathcal{Q}}\}$ in  (\ref{constraint5}) and  (\ref{constraint6}).

Then we prove the lower bound of the downlink communication. For $i\in[H]$, we define
\begin{subequations}
\begin{IEEEeqnarray}{rcl}
&&V_i=\left(\mathbf{v}_k: k \in\mathcal{N}_i\right), V_i^c=\left(\mathbf{v}_k: k \notin \mathcal{N}_i\right).
\end{IEEEeqnarray}
\end{subequations}
We have  
\begin{IEEEeqnarray}{rcl}
    H(X_0)&&\ge H(X_0|V_i)\nonumber\\
    &&=H(X_0|V_i, V_i^c)+I(X_0; V_i^c|V_i)\nonumber\\
    &&=H(X_0|V_i, V_i^c)+H(V_i^c|V_i)-H(V_i^c|V_i,X_0)\overset{(a)}{=}H(V_i^c|V_i)\overset{(b)}{=} H(V_i^c)\nonumber\\
    &&=(K-|\mathcal{N}_i|)\cdot V,
\end{IEEEeqnarray}
where $(a)$ holds because $H(X_0|V_i, V_i^c)=H(V_i^c|V_i,X_0)=0$, and $(b)$ is due to the assumption that $\mathbf{v}_k$ are i.i.d. random variables for $k\in[K]$. In some cases where the assumption can not hold due to the correlation between models, we can achieve the i.i.d. assumption by the encoding process before the uplink communication.

According to the definition of communication loads, we have 
\begin{IEEEeqnarray}{rcl}
 L_\textnormal{down}^{S\rightarrow \mathcal{E}*} &&\ge \frac{H(X_0)}{V}=K-|\mathcal{N}_i|\ge {K}-\min_{i\in[H]}|\mathcal{N}_i|\label{formula_fix_down}.
\end{IEEEeqnarray}
Combining (\ref{formula_fix_up}) and (\ref{formula_fix_down}), we complete the proof of the converse part of Theorem \ref{thm: lower}.

\subsection{Proof of Corollary \ref{corollary: gap}}\label{sec: gap}
In this subsection, we prove the gap between the achievable communication load in Theorem \ref{thm: upper} and the lower bound in Theorem \ref{thm: lower}. 

For the uplink communication of $z$ round, we can simply divide each IV $\mathbf{v}_k$ with $k\in \mathcal{I}_z$ into $z$ equal disjoint segments, and let relays send $n_i^z$ independent linear combinations of the segments following the schemes in (\ref{equ: uplinkScheme}).
Using this method, we have $\alpha^z_i =\frac{1}{z}$ if $|\mathcal{N}_i \cap \mathcal{I}_z| \neq 0$ and $\alpha^z_i = 0$ if $|\mathcal{N}_i \cap \mathcal{I}_z| = 0$, and it is obvious that the $\{\alpha^z_i\}_{i=1}^H$ is a feasible solution of the optimization problem $\mathcal{P}(z)$ as it meets the constraint of  in (\ref{equ: constraint1}) and (\ref{equ: constraint2}). According to (\ref{equ: up}),  the uplink communication load using equal division is
\begin{IEEEeqnarray}{rcl} \label{equ: upEqual}
    \hat{L}_{\textnormal{up}}^{\mathcal{E}\rightarrow S} &&= 
    \sum_{z=1}^{H-1} \sum_{i=1}^H  \alpha^z_i \left(
    \left| \mathcal{N}_i \cap \mathcal{I}_z  \right|  -\min_{j\in[H]} \left| \mathcal{N}_i \cap \mathcal{N}_j \cap \mathcal{I}_z  \right| \right),\nonumber\\
    &&=\sum_{z=1}^{H-1} \sum_{i=1}^H  \frac{1}{z} \left(
    \left| \mathcal{N}_i \cap \mathcal{I}_z  \right|  -\min_{j\in[H]} \left| \mathcal{N}_i \cap \mathcal{N}_j \cap \mathcal{I}_z  \right| \right),\nonumber\\
    &&=K- \sum_{z=1}^{H-1} \sum_{i=1}^H  \frac{1}{z} \min_{j\in[H]} \left| \mathcal{N}_i \cap \mathcal{N}_j \cap \mathcal{I}_z  \right|,
\end{IEEEeqnarray}
and we have $\hat{L}_{\textnormal{up}}^{\mathcal{E}\rightarrow S}\ge L_{\textnormal{up}}^{\mathcal{E}\rightarrow S}$ as $L_{\textnormal{up}}^{\mathcal{E}\rightarrow S}$ is achieved based on the optimal $\{\alpha^z_i\}_{i=1}^H$. 

For the uplink communication, from (\ref{equ: upEqual}) and (\ref{equ: downLower}), we have that
\begin{IEEEeqnarray}{rcl}
    L_{\textnormal{up}}^{\mathcal{E}\rightarrow S} - L_{\textnormal{up}}^{\mathcal{E}\rightarrow S*} && \le
    \hat{L}_{\textnormal{up}}^{\mathcal{E}\rightarrow S}-L_{\textnormal{up}}^{\mathcal{E}\rightarrow S*}\nonumber \\
    &&\le K- \sum_{z=1}^{H-1} \sum_{i=1}^H  \frac{1}{z} \min_{j\in[H]} \left| \mathcal{N}_i \cap \mathcal{N}_j \cap \mathcal{I}_z \right| - L_{\textnormal{up}}^{\mathcal{E}\rightarrow S*}  \nonumber \\ 
    &&\overset{(a)}{\le} K- \sum_{z=1}^{H-1} \sum_{i=1}^H  \frac{1}{z} \min_{j\in[H]} \left| \mathcal{N}_i \cap \mathcal{N}_j \cap \mathcal{I}_z \right| - L_{\textnormal{down*}}^{S\rightarrow \mathcal{E}} \nonumber\\
    && \le K- \sum_{z=1}^{H-1} \sum_{i=1}^H  \frac{1}{z} \min_{j\in[H]} \left| \mathcal{N}_i \cap \mathcal{N}_j \cap \mathcal{I}_z \right| - K \nonumber  +  \min_{i\in[H]}|\mathcal{N}_i| \nonumber \\
    && = \min_{i\in[H]}|\mathcal{N}_i| - \sum_{z=1}^{H-1} \sum_{i=1}^H  \frac{1}{z} \min_{j\in[H]} \left| \mathcal{N}_i \cap \mathcal{N}_j \cap \mathcal{I}_z \right|.
\end{IEEEeqnarray}
where $(a)$ is due to the fact that more bits need to
be transmitted in the uplink than in the downlink as the server knows all the messages sent from the uplink.

For the downlink communication, from  (\ref{equ: down}) and (\ref{equ: downLower}), we have that
\begin{IEEEeqnarray}{rcl}
    L_{\textnormal{down}}^{S\rightarrow \mathcal{E}} - L_{\textnormal{down*}}^{S\rightarrow \mathcal{E}} &&\le K-\sum_{z=1}^{H-1} \min_{i\in [H]}\left|\mathcal{N}_i \cap I_z\right| - {K}   +  \min_{i\in[H]}|\mathcal{N}_i|  \\
    &&= \min_{i\in[H]} |\mathcal{N}_i| - \sum_{z=1}^{H-1} \min_{i\in [H]}\left|\mathcal{N}_i \cap I_z\right|.  
\end{IEEEeqnarray}

\section{Proof of  Theorem \ref{corallary: flexible}}\label{sec: flex}
Consider a hierarchical distributed MTL system that consists of $K$ users and $H$ relays with average degree $r$, and we can delicately design the network connection. We consider a sufficiently large number of users $K$, where ${K}/{\binom{H}{r}}\in \mathbb{N}^+$ \footnote{If the number of users $K$ does not satisfy ${K}/{\binom{H}{r}}\in \mathbb{N}^+$, we first add some virtual users so that the condition can be met.}.

For the design of the network connection, we use the symmetric scheme as shown in \cite{li2017fundamental}. We first partition $K$ users into $\binom{H}{r}$ even disjoint groups of size $\eta = {K}/{\binom{H}{r}}$.  We denote each group as $\mathcal{K}_\mathcal{T}$, which corresponds to a unique set $\mathcal{T}\in[K]$ of size $|\mathcal{T}|=r$, i.e., $\{1,\ldots,K\}=\cup_{\mathcal{T}\in[K], |\mathcal{T}|=r} \mathcal{K}_{\mathcal{T}}$. The relay $i\in[H]$, connects to all the users in the group $\mathcal{K}_\mathcal{T}$ if $i\in \mathcal{T}$. Hence each relay $i$ connects to $|\mathcal{N}_i|=\binom{H-1}{r-1} \eta = \frac{rK}{H}$ users since each relay $i$ is in $\binom{H-1}{r-1}$ subset $\mathcal{T}$ of size $r$. 

After the local update phase and the uplink communication from users to relays, the relay gets the set of IVs $\{\mathbf{v}_k: k \in \mathcal{N}_i\}$. 
For any given relay, the common number of IVs shared by another relay is $\frac{H-2}{r-2}\eta = \frac{rK(r-1)}{H(H-1)}$, as any two relays are in $\binom{H-2}{r-2}$ subset $\mathcal{T}$ of size $r$. Recall $n_i^r$ denotes the maximum number of IVs available at relay $i$ but unavailable at other relay $j \in [H]\backslash\{i\}$, and it is obvious that all $n_i^r$, $i\in[H]$ are equal to $n^r$, where
\begin{IEEEeqnarray}{rcl}
    n^r=\frac{rK}{H}- \frac{rK(r-1)}{H(H-1)} = \frac{rK(H-r)}{H(H-1)}.
\end{IEEEeqnarray}

As each IV is available at $r$ relays, we only consider the $z=r$ communication round in Section \ref{sec: scheme}, and $\mathcal{I}_r=K$.
And the optimization problem in (\ref{equ: optimization_problem}) reduces to
\begin{subequations}
\begin{IEEEeqnarray}{rcl} 
    \mathcal{P}(r): \label{equ: optimization_problem_flex}\\
     \min_{\{\alpha^r_i\}_{i\in [H]}} && \sum_{i=1}^H n^r \alpha^r_i\\
    s.t.&&  \sum_{i=1}^H \alpha^r_i \mathds{1}_{\mathcal{N}_i}(k) \ge 1,  \forall k \in [K], \\ 
    &&  0 \le \alpha^r_i\le 1, \forall i \in [H].
\end{IEEEeqnarray}
\end{subequations}

Solving the above optimization problem, we have ${\{\alpha^r_i\}_{i\in [H]}}=\frac{1}{r}$. Based on Theorem \ref{thm: upper}, the uplink communication load $L_{\textnormal{up}}^{\mathcal{E}\rightarrow S}$  is
\begin{IEEEeqnarray}{rcl}
    L_{\textnormal{up}}^{\mathcal{E}\rightarrow S}&&=\sum_{i=1}^H n^r \alpha^r_i\\
    &&= H \cdot \frac{r K(H-r)}{H(H-1)} \cdot \frac{1}{r} = \frac{ K(H-r)}{H-1}.\label{equ: uplink_flex_upper}
\end{IEEEeqnarray} 

Using Theorem \ref{thm: upper}, the downlink communication load $L_{\textnormal{down}}^{S\rightarrow \mathcal{E}}$ is 
\begin{IEEEeqnarray}{rcl}
    L_{\textnormal{down}}^{S\rightarrow \mathcal{E}}&&= K- \min_{i\in[H]} |\mathcal{I}_r\cap \mathcal{N}_i | \\
    &&= K-\frac{rK}{H}.\label{equ: downlink_flex_upper}
\end{IEEEeqnarray}
Hence the communication load pair  $\left(\frac{ K(H-r)}{H-1}, K-\frac{rK}{H}\right)$ is achievable with the average degree $r$ under delicate design.

Substituting $\mathcal{I}_r$ and $|\mathcal{N}_i|$ in  (\ref{equ: upLower}) and  (\ref{equ: downLower}), we have that,
\begin{subequations}
\label{eqUpperFix}
\begin{IEEEeqnarray}{rCl} 
L_\textnormal{up*}^{\mathcal{E}\rightarrow S} &&\geq \frac{HK}{H-1}-\frac{rK}{H-1}=\frac{ K(H-r)}{(H-1)}, \\ 
L_\textnormal{down*}^{S\rightarrow \mathcal{E}} &&\geq {K}-\min_{i\in[H]}|\mathcal{N}_i|= K-\frac{rK}{H}.
\end{IEEEeqnarray}
\end{subequations}

Comparing the upper bounds in (\ref{equ: uplink_flex_upper}) and (\ref{equ: downlink_flex_upper}) with the lower bounds in \eqref{eqUpperFix}, we obtain that the proposed codes scheme is  optimal.

\bibliographystyle{IEEEtran.bst}
\bibliography{ref.bib}

% Generated by IEEEtran.bst, version: 1.14 (2015/08/26)
\begin{thebibliography}{10}
\providecommand{\url}[1]{#1}
\csname url@samestyle\endcsname
\providecommand{\newblock}{\relax}
\providecommand{\bibinfo}[2]{#2}
\providecommand{\BIBentrySTDinterwordspacing}{\spaceskip=0pt\relax}
\providecommand{\BIBentryALTinterwordstretchfactor}{4}
\providecommand{\BIBentryALTinterwordspacing}{\spaceskip=\fontdimen2\font plus
\BIBentryALTinterwordstretchfactor\fontdimen3\font minus
  \fontdimen4\font\relax}
\providecommand{\BIBforeignlanguage}[2]{{%
\expandafter\ifx\csname l@#1\endcsname\relax
\typeout{** WARNING: IEEEtran.bst: No hyphenation pattern has been}%
\typeout{** loaded for the language `#1'. Using the pattern for}%
\typeout{** the default language instead.}%
\else
\language=\csname l@#1\endcsname
\fi
#2}}
\providecommand{\BIBdecl}{\relax}
\BIBdecl

\bibitem{letaief2021edge}
K.~B. Letaief, Y.~Shi, J.~Lu, and J.~Lu, ``Edge artificial intelligence for 6g:
  Vision, enabling technologies, and applications,'' \emph{IEEE Journal on
  Selected Areas in Communications}, vol.~40, no.~1, pp. 5--36, 2021.

\bibitem{zhang2021survey}
Y.~Zhang and Q.~Yang, ``A survey on multi-task learning,'' \emph{IEEE
  Transactions on Knowledge and Data Engineering}, 2021.

\bibitem{ruder2017overview}
S.~Ruder, ``An overview of multi-task learning in deep neural networks,''
  \emph{arXiv preprint arXiv:1706.05098}, 2017.

\bibitem{tan2022towards}
A.~Z. Tan, H.~Yu, L.~Cui, and Q.~Yang, ``Towards personalized federated
  learning,'' \emph{IEEE Transactions on Neural Networks and Learning Systems},
  2022.

\bibitem{liu2017distributed}
S.~Liu, S.~J. Pan, and Q.~Ho, ``Distributed multi-task relationship learning,''
  in \emph{Proceedings of the 23rd ACM SIGKDD International Conference on
  Knowledge Discovery and Data Mining}, 2017, pp. 937--946.

\bibitem{jaggi2014communication}
M.~Jaggi, V.~Smith, M.~Tak{\'a}c, J.~Terhorst, S.~Krishnan, T.~Hofmann, and
  M.~I. Jordan, ``Communication-efficient distributed dual coordinate ascent,''
  \emph{Advances in neural information processing systems}, vol.~27, 2014.

\bibitem{smith2017federated}
V.~Smith, C.-K. Chiang, M.~Sanjabi, and A.~Talwalkar, ``Federated multi-task
  learning,'' in \emph{Proceedings of the 31st International Conference on
  Neural Information Processing Systems}, 2017, pp. 4427--4437.

\bibitem{dinh2021new}
C.~T. Dinh, T.~T. Vu, N.~H. Tran, M.~N. Dao, and H.~Zhang, ``A new look and
  convergence rate of federated multi-task learning with laplacian
  regularization,'' \emph{arXiv e-prints}, pp. arXiv--2102, 2021.

\bibitem{shi2020communication}
Y.~Shi, K.~Yang, T.~Jiang, J.~Zhang, and K.~B. Letaief,
  ``Communication-efficient edge ai: Algorithms and systems,'' \emph{IEEE
  Communications Surveys \& Tutorials}, vol.~22, no.~4, pp. 2167--2191, 2020.

\bibitem{mao2017survey}
Y.~Mao, C.~You, J.~Zhang, K.~Huang, and K.~B. Letaief, ``A survey on mobile
  edge computing: The communication perspective,'' \emph{IEEE communications
  surveys \& tutorials}, vol.~19, no.~4, pp. 2322--2358, 2017.

\bibitem{liu2020client}
L.~Liu, J.~Zhang, S.~Song, and K.~B. Letaief, ``Client-edge-cloud hierarchical
  federated learning,'' in \emph{ICC 2020-2020 IEEE International Conference on
  Communications (ICC)}.\hskip 1em plus 0.5em minus 0.4em\relax IEEE, 2020, pp.
  1--6.

\bibitem{prakash2020hierarchical}
S.~Prakash, A.~Reisizadeh, R.~Pedarsani, and A.~S. Avestimehr, ``Hierarchical
  coded gradient aggregation for learning at the edge,'' in \emph{2020 IEEE
  International Symposium on Information Theory (ISIT)}.\hskip 1em plus 0.5em
  minus 0.4em\relax IEEE, 2020, pp. 2616--2621.

\bibitem{sasidharan2022coded}
B.~Sasidharan and A.~Thomas, ``Coded gradient aggregation: A tradeoff between
  communication costs at edge nodes and at helper nodes,'' \emph{IEEE Journal
  on Selected Areas in Communications}, vol.~40, no.~3, pp. 761--772, 2022.

\bibitem{tse2005fundamentals}
D.~Tse and P.~Viswanath, \emph{Fundamentals of wireless communication}.\hskip
  1em plus 0.5em minus 0.4em\relax Cambridge university press, 2005.

\bibitem{ngai2004network}
C.~K. Ngai and R.~W. Yeung, ``Network coding gain of combination networks,'' in
  \emph{Information Theory Workshop}.\hskip 1em plus 0.5em minus 0.4em\relax
  IEEE, 2004, pp. 283--287.

\bibitem{zewail2018combination}
A.~A. Zewail and A.~Yener, ``Combination networks with or without secrecy
  constraints: The impact of caching relays,'' \emph{IEEE Journal on Selected
  Areas in Communications}, vol.~36, no.~6, pp. 1140--1152, 2018.

\bibitem{li2017fundamental}
S.~Li, M.~A. Maddah-Ali, Q.~Yu, and A.~S. Avestimehr, ``A fundamental tradeoff
  between computation and communication in distributed computing,'' \emph{IEEE
  Transactions on Information Theory}, vol.~64, no.~1, pp. 109--128, 2017.

\bibitem{tang2021communication}
H.~Tang, H.~Hu, K.~Yuan, and Y.~Wu, ``Communication-efficient coded distributed
  multi-task learning,'' in \emph{2021 IEEE Global Communications Conference
  (GLOBECOM)}.\hskip 1em plus 0.5em minus 0.4em\relax IEEE, 2021, pp. 1--6.

\bibitem{prakash2020coded}
S.~Prakash, S.~Dhakal, M.~R. Akdeniz, Y.~Yona, S.~Talwar, S.~Avestimehr, and
  N.~Himayat, ``Coded computing for low-latency federated learning over
  wireless edge networks,'' \emph{IEEE Journal on Selected Areas in
  Communications}, vol.~39, no.~1, pp. 233--250, 2020.

\bibitem{kairouz2021advances}
P.~Kairouz, H.~B. McMahan, B.~Avent, A.~Bellet, M.~Bennis, A.~N. Bhagoji,
  K.~Bonawitz, Z.~Charles, G.~Cormode, R.~Cummings \emph{et~al.}, ``Advances
  and open problems in federated learning,'' \emph{Foundations and
  Trends{\textregistered} in Machine Learning}, vol.~14, no. 1--2, pp. 1--210,
  2021.

\bibitem{maddah2014fundamental}
M.~A. Maddah-Ali and U.~Niesen, ``Fundamental limits of caching,'' \emph{IEEE
  Transactions on information theory}, vol.~60, no.~5, pp. 2856--2867, 2014.

\bibitem{karamchandani2016hierarchical}
N.~Karamchandani, U.~Niesen, M.~A. Maddah-Ali, and S.~N. Diggavi,
  ``Hierarchical coded caching,'' \emph{IEEE Transactions on Information
  Theory}, vol.~62, no.~6, pp. 3212--3229, 2016.

\bibitem{wang2019reduce}
K.~Wang, Y.~Wu, J.~Chen, and H.~Yin, ``Reduce transmission delay for
  caching-aided two-layer networks,'' in \emph{2019 IEEE International
  Symposium on Information Theory (ISIT)}.\hskip 1em plus 0.5em minus
  0.4em\relax IEEE, 2019.

\bibitem{ji2015fundamental}
M.~Ji, M.~F. Wong, A.~M. Tulino, J.~Llorca, G.~Caire, M.~Effros, and
  M.~Langberg, ``On the fundamental limits of caching in combination
  networks,'' in \emph{2015 IEEE 16th International Workshop on Signal
  Processing Advances in Wireless Communications (SPAWC)}.\hskip 1em plus 0.5em
  minus 0.4em\relax IEEE, 2015, pp. 695--699.

\bibitem{tang2016coded}
L.~Tang and A.~Ramamoorthy, ``Coded caching for networks with the resolvability
  property,'' in \emph{2016 IEEE International Symposium on Information Theory
  (ISIT)}.\hskip 1em plus 0.5em minus 0.4em\relax IEEE, 2016, pp. 420--424.

\bibitem{zewail2017coded}
A.~A. Zewail and A.~Yener, ``Coded caching for combination networks with
  cache-aided relays,'' in \emph{2017 IEEE International Symposium on
  Information Theory (ISIT)}.\hskip 1em plus 0.5em minus 0.4em\relax IEEE,
  2017, pp. 2433--2437.

\bibitem{zewail2019cache}
------, ``Cache-aided combination networks with asymmetric end users,'' in
  \emph{2019 IEEE 20th International Workshop on Signal Processing Advances in
  Wireless Communications (SPAWC)}.\hskip 1em plus 0.5em minus 0.4em\relax
  IEEE, 2019, pp. 1--5.

\bibitem{zhang2010convex}
Y.~Zhang and D.~Y. Yeung, ``A convex formulation for learning task
  relationships in multi-task learning,'' in \emph{Proceedings of the 26th
  Conference on Uncertainty in Artificial Intelligence, UAI 2010}, 2010, p.
  733.

\bibitem{zhou2011clustered}
J.~Zhou, J.~Chen, and J.~Ye, ``Clustered multi-task learning via alternating
  structure optimization,'' \emph{Advances in neural information processing
  systems}, vol. 2011, p. 702, 2011.

\bibitem{evgeniou2004regularized}
T.~Evgeniou and M.~Pontil, ``Regularized multi--task learning,'' in
  \emph{Proceedings of the tenth ACM SIGKDD international conference on
  Knowledge discovery and data mining}, 2004, pp. 109--117.

\bibitem{jacob2008clustered}
L.~Jacob, J.-p. Vert, and F.~Bach, ``Clustered multi-task learning: A convex
  formulation,'' \emph{Advances in Neural Information Processing Systems},
  vol.~21, pp. 745--752, 2008.

\bibitem{biggs1993algebraic}
N.~Biggs, N.~L. Biggs, and B.~Norman, \emph{Algebraic graph theory}.\hskip 1em
  plus 0.5em minus 0.4em\relax Cambridge university press, 1993, no.~67.

\bibitem{huang2013depth}
J.~Huang, F.~Qian, Y.~Guo, Y.~Zhou, Q.~Xu, Z.~M. Mao, S.~Sen, and
  O.~Spatscheck, ``An in-depth study of lte: Effect of network protocol and
  application behavior on performance,'' \emph{ACM SIGCOMM Computer
  Communication Review}, vol.~43, no.~4, pp. 363--374, 2013.

\bibitem{boyd2004convex}
S.~Boyd, S.~P. Boyd, and L.~Vandenberghe, \emph{Convex optimization}.\hskip 1em
  plus 0.5em minus 0.4em\relax Cambridge university press, 2004.

\bibitem{vaidya1987algorithm}
P.~M. Vaidya, ``An algorithm for linear programming which requires o (((m+ n) n
  2+(m+ n) 1.5 n) l) arithmetic operations,'' in \emph{Proceedings of the
  nineteenth annual ACM symposium on Theory of computing}, 1987, pp. 29--38.

\bibitem{aji2017sparse}
A.~F. Aji and K.~Heafield, ``Sparse communication for distributed gradient
  descent,'' in \emph{Proceedings of the 2017 Conference on Empirical Methods
  in Natural Language Processing}, 2017.

\bibitem{stich2018sparsified}
S.~U. Stich, J.-B. Cordonnier, and M.~Jaggi, ``Sparsified sgd with memory,'' in
  \emph{Proceedings of the 32nd International Conference on Neural Information
  Processing Systems}, ser. NIPS'18.\hskip 1em plus 0.5em minus 0.4em\relax Red
  Hook, NY, USA: Curran Associates Inc., 2018, p. 4452–4463.

\bibitem{wangni2017gradient}
J.~Wangni, J.~Wang, J.~Liu, and T.~Zhang, ``Gradient sparsification for
  communication-efficient distributed optimization,'' in \emph{Proceedings of
  the 32nd International Conference on Neural Information Processing Systems},
  ser. NIPS'18.\hskip 1em plus 0.5em minus 0.4em\relax Red Hook, NY, USA:
  Curran Associates Inc., 2018, p. 1306–1316.

\bibitem{alistarh2017qsgd}
D.~Alistarh, D.~Grubic, J.~Li, R.~Tomioka, and M.~Vojnovic, ``Qsgd:
  Communication-efficient sgd via gradient quantization and encoding,''
  \emph{Advances in Neural Information Processing Systems}, vol.~30, pp.
  1709--1720, 2017.

\bibitem{bernstein2018signsgd}
J.~Bernstein, Y.-X. Wang, K.~Azizzadenesheli, and A.~Anandkumar, ``signsgd:
  Compressed optimisation for non-convex problems,'' in \emph{International
  Conference on Machine Learning}.\hskip 1em plus 0.5em minus 0.4em\relax PMLR,
  2018, pp. 560--569.

\bibitem{liang2021improved}
K.~Liang and Y.~Wu, ``Improved communication efficiency for distributed mean
  estimation with side information,'' in \emph{{IEEE} International Symposium
  on Information Theory, {ISIT} 2021, Melbourne, Australia, July 12-20, 2021},
  2021, pp. 3185--3190.

\bibitem{grant2014cvx}
M.~Grant and S.~Boyd, ``Cvx: Matlab software for disciplined convex
  programming, version 2.1,'' 2014.

\bibitem{yu2017optimally}
Q.~Yu, S.~Li, M.~A. Maddah-Ali, and A.~S. Avestimehr, ``How to optimally
  allocate resources for coded distributed computing?'' in \emph{2017 IEEE
  International Conference on Communications (ICC)}.\hskip 1em plus 0.5em minus
  0.4em\relax IEEE, 2017, pp. 1--7.

\end{thebibliography}

\end{document}